\documentclass[11pt,a4paper]{article} 
\pdfoutput=1
\usepackage[no-natbib-sort]{my-jheppub}

\usepackage{amsmath, amssymb}
\usepackage{mathpazo}
\usepackage{environ}
\usepackage{mathrsfs}
\usepackage{array,arydshln}

\usepackage{graphicx,epsfig}
\usepackage{epic}
\usepackage{youngtab}
\usepackage{float}
\usepackage{color}
\definecolor{maroon}{rgb}{0.8,0.3,0.}

\usepackage{slashed}

\usepackage{hyperref}

\usepackage{aurical}
\usepackage[T1]{fontenc}

\newcommand{\be}{\begin{equation}}
\newcommand{\ee}{\end{equation}}

\newcommand{\ads}[1]{\text{AdS$_{#1}$}}

\newcommand{\mc}{\mathcal }
\newcommand{\mb}{\mathbb }

\newcommand{\Z}{\mathcal{Z}}

\newcommand{\M}{{\mathcal M}}

\newcommand{\bv}{\begin{verbatim}}
\newcommand{\ev}{\end{verbatim}}

\def \del{ \partial}
\def \la {\label}
\newcommand{\rf}[1]{(\ref{#1})}
\def\ov{\over}
\def\no{\nonumber} 
\def \ci {\cite}
\def \p {\phi}
\def \m {\mu}\def \n {\nu} 
\def \ed {\end{document}}

\def \foot {\footnote}
\def \b {\beta}

\def \D {\Delta} 
\def \vp {\varphi} 
 \def \ha {{{1 \ov 2}}}

\def \OO  {{\mc O}}      
 
\def \te {\textstyle} \def \iffa {\iffalse} 
\def \ha {{\te {1 \ov 2}}}
\def  \ba { \begin{align} }
\def  \ea { \end{align} }

\def \ep {\epsilon}
  
\def \RR {{\rm R}}
\def \OO {{\cal O}} 
\def \edd {\end{document}} 
\def \td {\tilde}

\def\s {\text{s}}
\def \vp {\varphi}

\def \iffa  {\iffalse}
\def \cp {\widehat {\p}}
\def \wCF {{\widehat{\rm CF}}}
\def \RR  {{\mathbb R}}
\def \s {\sigma}

\title{Iterating  free-field AdS/CFT:\\
 higher spin partition function relations
}
\author[a]{Matteo Beccaria\footnote{matteo.beccaria@le.infn.it}} 
\author[b]{ and \ \ Arkady A. Tseytlin\footnote{Also at Lebedev Institute, Moscow. \ \ tseytlin@imperial.ac.uk}}

\abstract{
We  find a  simple relation   between a free higher spin partition function on thermal quotient   of  \ads{d+1} 
and the partition function of the  associated  $d$-dimensional    conformal higher  spin  field  defined on 
 thermal quotient   of  \ads{d}. Starting with a conformal  higher  spin  field  defined in \ads{d} one may  also associate to it another  conformal field  in $d-1$  
  dimensions, thus "iterating" AdS/CFT.
 We  observe  that in  the case of $d=4$  this  iteration  leads to a  "trivial"  3d  higher spin conformal 
 theory   with  parity-even non-local action:  it  describes  zero total number of dynamical degrees  of freedom  and   the corresponding 
    partition   function 
    is  equal to 1. 
}

\affiliation[a]{Dipartimento di Matematica e Fisica Ennio De Giorgi,\\
Universit\`a del Salento \& INFN, Via Arnesano, 73100 Lecce, 
Italy} 

\affiliation[b]{The Blackett Laboratory, Imperial College, London SW7 2AZ, U.K.}

\allowdisplaybreaks

\begin{document}
\date{\currenttime}


 \begin{flushright}\small{Imperial-TP-AT-2016-{01}}\end{flushright}				

\maketitle
\flushbottom
\def \bea {\begin{eqnarray}}
\def \eea {\end{eqnarray}}
\def \D {\Delta} \def \M {{\cal M}}\def \g {\gamma} 
\def \addd {AdS$_{d+1}$}
\def \add {AdS$_{d}$}
\def \tp {{\widetilde \p}}

\def \bZ  {{\widetilde {\Z}}}

 \def \HS  {{\rm HS}}
 \def \CF  {{\rm CF}}
 \def \te {\textstyle} 
\def \fo {{\te {1\ov 4}}}
\def \ha {{\te {1\ov 2}}}
\def \cs {cs}

\def \ed {\end{document}}

\section{Introduction }

"Kinematical"    AdS/CFT   correspondence relates a    field  $\vp$ in  \ads{d+1} 
(e.g., with standard 2-derivative  action 
with some mass parameter $M^2$ or associated dimension $\D$) to a conformal field $\p$  at the boundary $\M^d = \del(\ads{d+1})$
with canonical dimension $\D^{-} = d- \D$. The value of  AdS mass  parameter   and thus of $\D^{-}$   determines 
the number of derivatives in the kinetic term  in the action for 
$\p$:   
\be  \la{01} S_d =\int d^d x \ \p\,  \del^k\,  \p \ , \ \ \qquad  \qquad k= d- 2 \D^{-} = 2\D  - d \ . \ee
For example, a  massless totally symmetric   higher spin field $\vp_s$ 
 in \ads{d+1}  is  associated   to a conformal higher spin  field  $\p_s$   with the action $S_d  =\int d^d x \ \p_s\,  P_s\,   \del^{2s+d-4}\,   \p_s$
 where $P_s$ is traceless transverse projector (see  \ci{Beccaria:2014jxa}  for a review  and refs.).
  From the  standard  AdS/CFT perspective  a massless AdS  field  $\vp_s$ is  a counterpart of  a bilinear  conserved  current  $J_s$ 
  of a free (e.g., scalar $\Phi$) boundary  CFT$_d$  while $\p_s$ is  associated  with a shadow   field  or a source for $J_s$; 
   thus 
  the action for $\p_s$   may be interpreted as  
   an "induced" action   found  upon   integrating out $\Phi$   coupled to $\p_s$ via $J_s(\Phi)$.

   There are   other  relations between the two   free  theories  $S_{d+1}(\vp)$ and $S_d(\p)$ 
    in $d+1$ and $d$   dimensions   beyond just  kinematic
     $SO(2,d)$  representation theory  correspondence.  
     First, 
the \ads{d+1}  action for $\vp$   evaluated  on the solution of the Dirichlet problem $\vp\big|_{\del} = \p$
gives  an   "induced"  action for $\p$.  For  even $d$  and, e.g.,  for a massless  field $\vp$   the  \ads{d+1}   action   contains   a logarithmically divergent 
 local  term  which is identified with a local action for $\p$. 
 For odd $d$ one gets,  in general,  a non-local action as  the power $k$ in  the   kinetic operator in \rf{01} 
may be half-integer or negative.  
 In addition to this   tree-level relation there is also a 1-loop one --   between  ratio of  partition functions  for 
 a higher spin (HS)  field $\vp$ in 
 \ads{d+1} with 
 D and N (or $+$ and $-$)  boundary conditions    and  for the   dual conformal field  (CF)   $\p$   at the boundary: 
 \be \la{1.1}
{ Z^- _\HS  \ov Z^+_\HS}\Big|_{\ads{d+1}  }  = Z_\CF \Big|_{\del ({\ads{d+1}  } )  } \ . 
\ee
This relation is true, e.g., for global   \ads{d+1}    with boundary $S^d$   and also  for  a  thermal quotient   of \ads{d+1} 
with  boundary $S^1_\b  \times S^{d-1}$. In the latter case    \rf{1.1}   translates into a relation    between  one-particle partition functions 
$\Z$ as functions of the $q= e^{-\b}$  variable\foot{Here $\sigma_{d} (q)$ is a finite polynomial in $q+q^{-1} $
that   represents    contribution  of  
finite number of  "zero" modes  related to gauge invariance of the conformal  (shadow) field  
\ci{Giombi:2013yva,Beccaria:2014jxa}.}
\bea
\label{1.2}
&&\Z^{-}_\HS (\ads{d+1}; q)-\Z^+_\HS (\ads{d+1}; q) = \Z_\CF (S^{1}\times S^{d-1}; q) \ ,  \\
\la{222}
&&Z =  \exp \sum_{n=1}^\infty  \te {1\ov n } \Z( q^n) \ , \\
&&\la{112} \Z^{-}(\ads{d+1}; q) = \bZ^-(\ads{d+1}; q) + \sigma_{ d+1} (q) \ , \ \ \ \ \ \ \\
&& \la{221} 
\bZ^-(\ads{d+1}; q) \equiv (-1)^d \Z^+(\ads{d+1}; q^{-1})  \  . 
\eea
Eq. \rf{1.2}    may be  interpreted    in terms of counting of  operators in the boundary CFT or    as a group-theoretic   relation for characters of the conformal group. 
More generally, \rf{1.1}  is expected to be true   even for asymptotically AdS space  and its generic curved boundary 
(provided  the corresponding $d+1$ and $d$ dimensional theories can be  consistently defined)  and  should 
thus provide, in particular, an AdS theory   based   way to compute   not only the  conformal anomaly  a-coefficients \ci{Giombi:2013yva} 
 but also the c-coefficients  \ci{Beccaria:2014xda}.

Having identified a  conformal field  $\p$   in $\RR^d$ associated to a field $\vp$ in \addd 
we may  attempt to  repeat this step one more time.  Namely,  we may  first define this 
$\p$  not on $\RR^d$ (or $S^d$ or $S^1 \times S^{d-1}$)     but  on \add  \ and then  associate to it {\it another}  conformal field $\cp$ 
in $d-1$ dimensions.   
We will then have  the following dimensional ($d+1\to d \to d-1$)  digression\footnote{For standard global \ads{d+1}  we have 
$\partial \ads{d+1} = \RR \times S^{d-1}$. This space 
is equivalent to two copies of $\ads{d}$ glued along their boundary  identified with the equator of $S^{d-1}$.
The middle step in (\ref{1.3}) means that we  start with  the conformal action on $\partial \ads{d+1}$ and  then 
translate it  into   $\ads{d}$ (taking also into account the freedom in  choice of boundary conditions, see below).}
\be \la{1.3}
\vp (\ads{d+1}) \ \  \to \ \   \p (\del\,\ads{d+1} \propto \ads{d} ) \ \  \to \ \   \cp (\del\,  \ads{d})  \ .  \ee
If   $\vp$  is  a gauge  field  with 2-derivative action in \ads{d+1},  then 
$\p$  is  also  single (gauge) conformal  field  with, in general,  
 higher derivative action. The latter   can  be represented as a collection of 2nd-derivative 
 fields in \add\ and   hence   $\cp$  in $d-1$ dimensions  will    be    given by set 
  of  several conformal fields, each  being dual to  an individual 2nd-derivative field   in  $d$ dimensions. 

Our aim below will be  to    explore some  implications of this "iterated"    AdS/CFT   correspondence 
 \rf{1.3} at the level of  relations between partition functions  like \rf{1.1}  and \rf{1.2}.  
 We shall find that  for  a generic    higher spin field (HS)  in \ads{d+1} 
 and its  dual  conformal field (CF) in $d$ dimensions  one gets also 
 \be
\label{313}
\Z^{-}_{\rm HS}(\ads{d+1}; q) - \Z^{+}_{\rm HS}(\ads{d+1}; q)  = 
\Z^{-}_{\text{CF}}(\ads{d}; q) +\Z^{+}_{\text{CF}}(\ads{d}; q) \ , 
\ee
where $\Z^{-}$   may be replaced by $\bZ^-$ in \rf{221}  as  one finds  also that the $\sigma$ terms 
in \rf{112}  match, $\sigma_{_{\HS,\, d+1}}= \sigma_{_{\CF,\, d}}$. 
Eq. \rf{313}    follows from \rf{1.2} and 
\be \la{123}
  \Z^{-}_{\text{CF}}(\ads{d}; q) +\Z^{+}_{\text{CF}}(\ads{d}; q)  =\Z_{\text{CF}}(S^1\times S^{d-1}; q) \ , 
 \ee 
 which may be related to the fact that $\ads{d}$   is conformal to half of $\RR \times S^{d-1}$ 
 so that the  respective  partition functions  are related   provided one  sums over the two possible boundary conditions
 at the boundary of \ads{d}. 
 Applying  \rf{1.2} to  CF in \ads{d}   and its  counterpart  conformal field $\wCF$  in $d-1$ dimensions (cf. \rf{1.3}) 
 we get also 
 \be \la{234}
  \Z^{-}_{\text{CF}}(\ads{d}; q) - \Z^{+}_{\text{CF}}(\ads{d}; q)  =\Z_{{\wCF}}(S^1\times S^{d-2}; q) \ . 
 \ee 
 We shall   find that the case of $d=4$ is special:   starting with a HS   field 
 in \ads{5}, the  resulting  3d  conformal theory  represented  by $\cp$ 
 is effectively  "topological",   having  zero number of  dynamical d.o.f. 
 and  trivial   partition function. This  may  be related to equivalence of 
 $\pm$    modes   with non-zero spins  in \ads{4} \ci{Breitenlohner:1982bm}, implying 
 \be \la{345}
  \Z^{-}_{\text{CF}}(\ads{4}; q) = \Z^{+}_{\text{CF}}(\ads{4}; q) \quad \to \quad 
  \Z_{{\wCF}}(S^1\times S^{2}; q)=0  \  . 
 \ee 
 Very    loosely,   this  may be  interpreted  as  a   version of the "boundary of boundary =0"  
 relation, or as  "$(\text{AdS/CFT})^2 =0$".\foot{Let us  note that our  interpretation  and examples   
 will be  different from previous discussions 
  of   "sequential"  AdS/CFT  like $\ads{4}/\text{CF}_{3}\to \ads{3}/\text{CF}_{2}$ 
   in \cite{Nilsson:2012ky,Ohl:2012bk}
 (for related work    discussing \ads{d} foliations  of \ads{d+1} see also \ci{Karch:2000ct,Compere:2008us,Aharony:2010ay,Andrade:2011nh,Hinterbichler:2015pta}). 
 In particular, in contrast to  \cite{Nilsson:2012ky}  the 3d conformal  higher spin theory  that 
will naturally appear in our  context     is  not of local Chern-Simons type  but  has   parity-even  non-local
action. 
 Let us   also mention 
 for completeness  that  
  discussions  of  dimensional  reduction from to AdS$_{d+1}$ to AdS$_d$ 
 appeared in \cite{Metsaev:2000qb,Artsukevich:2008vy}.}
  
We shall start in section 2  with a review of some  general  definitions   and relations.  
Then in  section 3 we shall demonstrate the validity   of \rf{3.1} on several   examples, 
in particular, for massless   higher spin  fields in \ads{d+1}   related to conformal higher spin fields in \ads{d}. 

In section 4  we shall  first  analyse  the detailed  structure of   the relation \rf{3.1} on the example 
of  the totally symmetric   field in \ads{5}  with generic  mass  parameter  and mention its possible group-theoretic interpretation 
and then justify the $\Z^-=\Z^+$  equality  in \rf{345}. We shall then discuss in detail  the  corresponding 3d 
conformal theory with non-local linearized action   describing total of zero degrees of  freedom and leading to 
trivial partition function. We shall  use  spin 1 Maxwell and spin 2 conformal graviton  fields as examples. 

Section 5 will  contain some concluding remarks. 
In Appendix \ref{a1}  we shall  discuss the  algebraic  structure of the partition functions  appearing in \rf{3.1} 
and  then in Appendix \ref{a2} argue   for the equality of the corresponding $\sigma$-terms in \rf{112}.
In Appendices \ref{a3}, \ref{a4} and \ref{a5}  the  relation \rf{3.1}   will be further  illustrated on the   examples 
of conformal   higher derivative scalars, fermionic conformal   higher spin fields and 
conformal  antisymmetric tensor field in 4d.


\section{Some general relations}

Let us consider a  conformally  invariant action in \ads{d}. This space is conformally equivalent to 
one half of static Einstein universe  $S^{1}\times S^{d-1}$, with the boundary of \ads{d} being mapped to the equator 
of $S^{d-1}$  \cite{Breitenlohner:1982bm,Dowker:1983nt,Hawking:1999dp}. One can consider 
the single particle partition function $\Z(\ads{d}; q)$ on thermal \ads{d} where we identify $t\sim t+\beta$.
This can be compared with the partition function in Einstein universe  $\Z(S^{1}\times S^{d-1}; q)$ where $S^{1}$ is the
thermal circle with length ${\beta}$.

The calculation of total partition function $Z(\ads{d}; q)$   (and thus of $\Z(\ads{d}; q)$)  is  straightforward
 assuming  that  the kinetic operator of a conformal field factorises, i.e. 
the action in \ads{d}   can be written a sum of 2nd-derivative terms (as, e.g., in \ci{Metsaev:2007rw}). 
For example, let us consider
\be
\label{2.1}
\log Z(\ads{d}) =-\ha \sum_{i=1}^{N} n_{i}\,\log\det\widehat \Delta_{s_{i}\perp}(M_{i}^{2})\ ,\qquad \qquad \widehat\Delta_{s\perp}(M^{2}) \equiv  (-\nabla^{2}-M^{2})_{s\perp}
\ee
where  $\widehat\Delta_{s\perp}$ is 
defined on transverse traceless  symmetric  tensors of rank $s$ \footnote{In general, we define  $\widehat\Delta_{s\perp}(M^{2}) =  (-\nabla^{2} + M^{2}\epsilon)_{s\perp}$, where 
$\epsilon=-1$ for \ads{d}  and $\epsilon=+1$ for $S^d$  (here we set the curvature scale to 1).},
and the integers $n_{i}$ are field multiplicities positive for physical fields and negative for ghost fields.
For each operator  in  \rf{2.1}  the  value of mass term then  determines  possible ground state
energies $\Delta^{\pm}_{d}$ that are solutions of the quadratic equation \cite{Metsaev:1994ys,Metsaev:1995re,Metsaev:2003cu}
\be
\label{2.2}
\Delta^{\pm}_{d}\,(\Delta^{\pm}_{d}-d+1)-s = -M^{2}, \qquad\qquad \ \ \  \Delta^{-}_{d} = d-1  -  \Delta^{+}_{d} \ , \ \quad   
 \Delta^{-}_{d}   \le  \Delta^{+}_{d} \ , 
\ee
 and are  associated with classical solutions of the wave equation 
$
\widehat\Delta_{s\perp}(M^{2})\,\vp_{s\perp}=0 
$
with two different boundary conditions. 
Taking the thermal quotient of \ads{d},  we then get from   (\ref{2.1})
the following two possibilities for the  corresponding 
single particle  partition function ($q= e^{-\b}$)
\be
\label{2.4}
\Z^{+}(\ads{d}; q) = \sum_{i=1}^{N} n_{i}\,g_{s_{i}}^{(d)}\,\frac{q^{\Delta^{+}_{d, i}}}{(1-q)^{d-1}}\  ,
\ \ \ \ \ \ \ \ 
\bZ^{-}(\ads{d}; q) = \sum_{i=1}^{N} n_{i}\,g_{s_{i}}^{(d)}\,\frac{q^{\Delta^{-}_{d, i}}}{(1-q)^{d-1}}\  .
\ee
In (\ref{2.4})  $g_{s}^{(d)}$ is the multiplicity
 that counts the number of off-shell degrees of freedom\footnote{Special cases are $g_{s}^{(4)} = 2s+1$, $g_{s}^{(6)} = \frac{1}{6}(s+1)(s+2)(2s+3)$.
}
\be
\label{2.5}
g_{s}^{(d)} = (2s+d-3)\frac{(s+d-4)!}{(d-3)!s!}.
\ee
Using that   $\Delta^{-} = d-1-\Delta^{+}$  we  find  
\be
\label{2.6}
\bZ^{-}(\ads{d}; q) = (-1)^{d-1}\,\Z^{+}(\ads{d}; q^{-1})\  .
\ee
In the presence  of gauge invariance  the proper $\Z^{-}(\ads{d}; q)$ partition function 
differs from $\bZ^{-}(\ads{d}; q)$  
\be \la{266}
  \Z^{-}_{\rm HS}(\ads{d}; q) = \bZ^{-}_{\rm HS}(\ads{d}; q)  + \sigma_{d}(q) \ , 
\ee
where 
$\sigma(q)$ is a finite polynomial in    $q+q^{-1}$ related to missing gauge transformations 
discussed in  \cite{Beccaria:2014jxa}.

The calculation of $\Z (S^{1}\times S^{d-1}; q)$ on the Einstein universe background  is a priori unrelated to the one on AdS$_{d}$.
If the action on generic ${\cal M}^d$   is known, one may just specialize it to $S^{1}\times S^{d-1}$, 
factorize the kinetic operator   and use the methods discussed in \cite{Beccaria:2014jxa}.\footnote{If one knows the set of masses $M^{2}$   in \rf{2.1}  for an  action on AdS$_{d}$, this is not enough 
 to compute the partition function for the same theory on $S^{1}\times S^{d-1}$. The reason is that $M^{2}$  values 
come from the specialization to AdS$_{d}$ of the action on a  generic curved background ${\cal M}^d$  where 
 certain combinations of  curvature tensor terms lead to mass terms. Specification   of this  action to $S^{1}\times S^{d-1}$
 will then lead to different  kinetic term structures, i.e. to different  mass terms in the corresponding 2nd-order operators. 
}
 Alternatively, one can make use of the  conformal  map to  flat space $\mb R^{d}$ ("radial quantization") 
 and use flat space operator counting techniques. 

At the same time, $\Z (S^{1}\times S^{d-1}; q)$    can also  be computed  starting with the  dual   theory in \ads{d+1} (cf. \rf{1.2}). 
If  $S^{1}\times S^{d-1}$ is interpreted as the boundary of \ads{d+1}  the  corresponding 
conformally  invariant action  on $S^{1}\times S^{d-1}$   can be interpreted as   
   "{induced}"  from an action  of a dual field in the bulk (see, e.g.,  \cite{Giombi:2013yva,Beccaria:2014jxa}  and refs. there). 
   Let us   call  a generic   tensor   bulk field a "higher spin" (HS)   one; this  name   will  include  the  
      cases of a  massive or partially  massless
or exactly massless higher spin fields in \ads{d+1}. 
The   dual conformal field at the boundary will be denoted as CF. 
Then    \cite{Beccaria:2014jxa}  
\be
\label{2.7}
\Z^{-}_{\rm HS}(\ads{d+1}; q)-\Z^{+}_{\rm HS}(\ads{d+1}; q) = \Z_{\rm CF}(S^{1}\times S^{d-1}; q)
\ . \ee
In the case of   massless    higher spin (MHS)  field in \ads{d+1}  having maximal   gauge invariance  the  associated 
 conformal field at the boundary is  conformal higher spin (CHS) one and $\sigma_{d+1}$ in \rf{266}  is non-trivial  \cite{Beccaria:2014jxa}.\foot{In addition to the quantum  "one-loop"   relation \rf{2.7}  the quadratic actions for  HS and CF  have also   classical relation: 
 evaluating the action  of HS field in \ads{d+1} on the solution  with boundary data  being equal to CF field 
 one gets the action of the CF  field as an "induced" one. In even $d$ case  the  local CF action is the coefficient of the leading logarithmic IR divergence while in odd  $d$ case it is finite but  non-local.}

\section{Partition functions on  \ads{d+1} and   \ads{d}}

Let us now   propose and check on 
 several examples  a general relation between partition 
functions  of    higher spin   field  in  \ads{d+1} and   associated   conformal  field  originally  "induced"   on 
  $\partial\ads{d+1} = \RR \times S^{d-1} $  or $S^d$   but that can then  be also   defined   on   \ads{d}.
%
This relation  is \rf{313} that we rewrite here for the reader's convenience
\iffa 
\footnote{Let us  remark that the Casimir energy associated with a generic partition function
$\Z(q) = \sum_{n}c_{n}\,q^{n} =  \sum_{n}c_{n}\,e^{-\beta\,n}$, 
is formally $E_{c} = \frac{1}{2}\,\sum_{n}n\,c_{n}$.
The usual way to regularize this quantity is to extract it from the Laurent expansion of $\Z(e^{-\beta})$
around $\beta=0$ , $\Z(e^{-\beta}) = \cdots -2\,E_{c}\,\beta + \mc O(\beta^{2})$,
where dots stand for poles and a constant. For a massive representation in AdS$_{d+1}$ one finds
$E_{c}(\Delta^{+}) = (-1)^{d+1}\,E_{c}(\Delta^{-})$.
This means that the Casimir energy from $\Delta^{+}$ and $\Delta^{-}$ is equal in the combinations
appearing in (\ref{3.1}).
}
\fi 
\be
\label{3.1}
\Z^{-}_{\rm HS}(\ads{d+1}; q) - \Z^{+}_{\rm HS}(\ads{d+1}; q)=\Z^{-}_{\text{CF}}(\ads{d}; q) +\Z^{+}_{\text{CF}}(\ads{d}; q)\ .
\ee
Heuristically, the  relation \rf{3.1} may be motivated as follows. 
$\ads{d}$ is conformal to  half of  the Einstein universe $S^{1}\times S^{d-1}$ with two possible choices of the boundary conditions
at the equator; thus  defining  the partition function on $S^{1}\times S^{d-1}$  in terms of \ads{d}  one
 we may need to sum over the two 
boundary condition choices, 
\be
\label{3.2}
\Z^{-}_{\text{CF}}(\ads{d}; q) +\Z^{+}_{\text{CF}}(\ads{d}; q)=\Z_{\rm CF}(S^{1}\times S^{d-1}; q)  \ .
\ee
Combining this with \rf{2.7} then gives \rf{3.1a}. 

Note that   starting with a CF field   in \ads{d}  we may also  associate to it another conformal field $\wCF$
at the $d-1$ boundary and then the analog of \rf{2.7}   will read 
\be
\label{3322}
\Z^{-}_{\rm CF}(\ads{d}; q)-\Z^{+}_{\rm CF}(\ads{d}; q) = \Z_{\wCF} (S^{1}\times S^{d-2}; q)
\ . \ee
Furthermore, the $\sigma(q)$   terms in \rf{266}  for  $\Z^{-}_{\rm HS}(\ads{d+1}; q)$ 
and $\Z^{-}_{\text{CF}}(\ads{d}; q)$ appear to match (see Appendix \ref{a2}) 
\be   \sigma_{_{{\rm HS}, \, {d+1}}}(q)=  \sigma_{_{{\rm CF}, \, {d}}}(q)  \  \la{388}\ee
so that \rf{3.1}   may be written also as 
\be
\label{3.1a}
\bZ^{-}_{\rm HS}(\ads{d+1}; q) - \Z^{+}_{\rm HS}(\ads{d+1}; q)  = 
\bZ^{-}_{\text{CF}}(\ads{d}; q) +\Z^{+}_{\text{CF}}(\ads{d}; q)\ .
\ee
Using \rf{2.6} 
this  can   be  put also  in the following more symmetric form 
\be
\label{3.8}
- \Z^{+}_{\rm HS}(\ads{d+1}; q) + (-1)^d  \Z^{+}_{\rm HS}(\ads{d+1}; q^{-1})  = 
 \Z^{+}_{\text  CF }(\ads{d}; q)  -  (-1)^d  \Z^{+}_{\text{CF}}(\ads{d}; q^{-1}) \quad
\ee
Below  we   will   demonstrate the validity of  \rf{3.1},\rf{3.1a}  and 
 \rf{3.2}  on   several  examples of conformal fields 
(for some  consequences of \rf{3.1a}  see   also   Appendix \ref{a1}).


\subsection{Conformal scalar}
\label{subsec:confsc}

Let us  start   with the  case of a  particular  scalar field  in \ads{d+1} with the 
mass term $M^2 =  \fo d^2 +1$, i.e. with 
$\Delta^{\pm}_{d+1} = \frac{1}{2}(d\pm 2)$  (cf. \rf{2.2}). The corresponding partition \rf{2.4} function is 
\be
\label{3.3}
\Z^{\pm}_{0}(\ads{d+1}; q) = \frac{q^{\frac{1}{2}(d\pm 2)}}{(1-q)^{d}}   \  .
\ee
This  scalar  in \ads{d+1}   "induces"   a  spin 0 field $\varphi$   at  the boundary
with   canonical dimension $=\Delta^{-}_{d+1} = \frac{1}{2}(d-2)$, i.e. 
  which thus   represents   
a conformally coupled scalar. The  corresponding   kinetic operator in a  
curved $d$-dimensional space   specified   to the case of the unit-scale  \ads{d} 
(with $R= - d (d-1)$ is 
\be  \la{3.4}\te 
- \nabla^2  +   { d-2 \ov 4 (d-1) }  R = - \nabla^2  - \fo d (d-2)    \ . 
\ee
Thus  defining  $\varphi$ 
on $\ads{d}$  
 we find  that the  mass term (cf. \rf{2.1}) is  $M^2=\fo d (d-2)$    and thus  from  \rf{2.2} 
\be
\label{3.5}\te 
\Delta_{d}^{+} = \frac{d}{2}, \qquad\qquad 
\Delta_{d}^{-} = \frac{d-2}{2}.
\ee
From \rf{2.4} the    partition functions  corresponding to this   conformal scalar  (cs) are then 
\be
\label{3.6}
\Z^{+}_{\rm cs}(\ads{d}; q) =  \frac{q^{d/2}}{(1-q)^{d-1}}, \qquad\qquad
\Z^{-}_{\rm  cs}(\ads{d}; q)  = (-1)^{d+1}\, \Z^{+}_{\rm cs}(\ads{d}; q^{-1})\ ,
\ee
Comparing to \rf{3.3}  one can  then check that 
\be
\label{3.7}
\Z^{-}_{0}(\ads{d+1}; q) - \Z^{+}_{0}(\ads{d+1}; q)  = 
\Z^{-}_{\text\cs}(\ads{d}; q) +\Z^{+}_{\text{cs}}(\ads{d}; q) \ ,
\ee
which is a particular  spin 0  case of \rf{3.1}.

To demonstrate \rf{3.2}   we recall that 
the partition function on $S^{1}\times S^{d-1}$ can   be 
 found, e.g.,  by the operator counting method. For  
a scalar  $\varphi$ with canonical dimension $\frac{1}{2}(d-2)$ 
and equations of motion $\partial^{2}\varphi=0$ that   gives
\be
\label{3.9}
\Z_{\text\cs}(S^{1}\times S^{d-1}; q) = \frac{q^{\frac{1}{2}(d-2)}-q^{\frac{1}{2}(d+2)}}{(1-q)^{d}}\ .
\ee
Then  one can check that \rf{2.7} is satisfied 
 (there is no gauge invariance  so  $\sigma(q)=0$ in \rf{266}).  As a result, we  verify   a special case of \rf{3.2}
\be
\label{3.10}
\Z_{\text\cs}(S^{1}\times S^{d-1}; q) = 
\Z^{-}_{\text\cs}(\ads{d}; q) +\Z^{+}_{\text\cs}(\ads{d}; q)\ .
\ee
As  already  mentioned above,  this relation means  that 
 one  needs to sum over 
 both $\pm$   scalar modes in \ads{d} in order to match the conformal scalar 
 partition function on  $S^{1}\times S^{d-1}$  space 
which is conformally equivalent to a double copy of \ads{d}.


The above discussion  can  be extended to higher derivative  GJMS conformal scalars
with higher-derivative  kinetic operators, see  Appendix \ref{a3}, 
again  verifying  the general relations \rf{3.1}--\rf{3.1a}.

\subsection{Conformal higher spins}

Let us now consider a totally   symmetric  spin $s$ CHS field  in  $d$ dimensions.
The  CHS theory  defined   on $AdS_{d}$  has the following  partition function \ci{Tseytlin:2013jya,Tseytlin:2013fca} 
\ba
&Z_{\text{CHS}, \, s}(\ads{d}) = \prod_{k=0}^{s-1}\Big[
\frac{\det\widehat\Delta_{k\perp}(M^2_{k,s})}
{\det\widehat\Delta_{s\perp}( M^2_{s,k})}
\Big]^{1/2}
\ \prod_{k'=-\frac{1}{2}(d-4)}^{-1}\Big[
\frac{1}{\det\widehat\Delta_{s\perp}(M^2_{s,k'})}
\Big]^{1/2}\ ,  \la{3.12} \\
& \qquad \qquad M^2_{n,k} \equiv    n - (k-1) (k+ d-2) \ .  \label{3.122}
\end{align}
For each determinant   here (cf. \rf{2.1})
 we may then compute the corresponding contribution to the one-particle partition function using the general relations \rf{2.2},\rf{2.4}. As a  result, we get 
\bea
&&\Z_{\text{CHS}, \, s}^{+}(\text{AdS}_{d}; q) = \frac{1}{(1-q)^{d-1}}\Big\{
\sum_{k=0}^{s-1} \Big[g_{s}^{(d)}\,q^{d+k-2}-g_{k}^{(d)}\,q^{d+s-2}\Big]
+\sum_{k'=-\frac{1}{2}(d-4)}^{-1} g_{s}^{(d)}\,q^{d+k'-2}
\Big\}\ ,  \no \\
&& \qquad \qquad \bZ_{\text{CHS}, \, s}^{-}(\text{AdS}_{d}; q) = 
(-1)^{d-1}\, \Z_{\text{CHS}, \, s}^{+}(\text{AdS}_{d}; q^{-1})\ .\label{3.13}
\eea
Doing the sum, we find 
\be
\begin{split}
\label{3.14}
\Z_{\text{CHS}, \, s}^{+}(\text{AdS}_{d}; q) &= \frac{\Gamma (d+s-3)}{\Gamma (d-1) \Gamma (s+1)}\,\frac{q^{\frac{d}{2}-2}}{(1-q)^{d}}\,
\Big[
(d-2)  (d+2 s-3) q^{2} \\
&
\qquad -(d+s-3) (d+2 s-2) q^{\frac{d}{2}+s}+s (d+2 s-4)
   q^{\frac{d}{2}+s+1}
   \Big]\ . 
   \end{split}
\ee
In the special case of the $s=2$  CHS field or, equivalently, of Weyl  gravity 
on  thermal quotient of \ads{4} 
and \ads{6}  this partition function was independently  computed  also   in \cite{Irakleidou:2015exd,Lovrekovic:2015thw} 
\be
\la{3144}\te 
\Z_\text{CHS,2}^{+}(\ads{4}; q) = \frac{q^{2}\,(5+5q-4q^{2})}{(1-q)^{3}}, \qquad \quad 
\Z_\text{CHS,2}^{+}(\ads{6}; q) = \frac{2q^{3}\,(7+7q+7q^{2}-3q^{3})}{(1-q)^{5}} \ .
\ee
The CHS field in $d$ dimensions 
 is naturally associated to  the  massless higher spin (MHS) field 
 in AdS$_{d+1}$ with $\Delta^{+}_{d+1} = d+s-2$.
  It has the  one-particle partition function  \ci{Gopakumar:2011qs,Gupta:2012he,Giombi:2014yra}
\bea
&&
\label{3.11}
\Z_{\text{MHS}, \, s}^{+}(\text{AdS}_{d+1}; q) = \frac{g_{s}^{(d+1)}\,q^{d+s-2}-g_{s-1}^{(d+1)}q^{s+d-1}}{(1-q)^{d}}\ , \\
&&\bZ_{\text{MHS}, \, s}^{-}(\text{AdS}_{d+1}; q) = 
(-1)^{d}\, \Z_{\text{MHS}, \, s}^{+}(\text{AdS}_{d+1}; q^{-1})  \ , 
\eea
where $g_{s}^{(d)}$ is given by \rf{2.5}. 
One can then  check that 
\be
\label{3.15}
\bZ_{\text{MHS}, \, s}^{-}(\text{AdS}_{d+1}; q) -\Z_{\text{MHS}, \, s}^{+}(\text{AdS}_{d+1}; q) = 
\bZ_{\text{CHS}, \, s}^{-}(\text{AdS}_{d}; q) +\Z_{\text{CHS}, \, s}^{+}(\text{AdS}_{d}; q) \ , 
\ee
which is another special case of \rf{3.1a}. 
One can also verify  the validity of \rf{3.1} or, equivalently,  \rf{388} (see  Appendix \ref{a2}).

\subsection{Conformal symmetric tensors in $d=4$}

Next, let us discuss  the conformal symmetric  rank $s$  tensor field (CST) in $d=4$   considered in \cite{Erdmenger:1997wy,Beccaria:2015vaa}.
This is a non-unitary theory that may be viewed as a maximal depth $r = s$ representative of the
family of FT-type  \ci{Fradkin:1985am}   conformal higher spin fields    with rank $s-r$  tensor gauge invariance 
\cite{Vasiliev:2009ck,Bekaert:2013zya,Barnich:2015tma}. 
The CHS theory is the minimal depth  case (i.e.  case of   maximal gauge invariance)    when $r = 1$. 
The CST   field  has  2nd-derivative Lagrangian with scalar gauge invariance 
and   corresponds to  a  "short" representation of $SO(2,4)$  given by 
\be \la{319}
{\rm CST}_s = \te (1;{s\ov 2}, {s\ov2}) - (1-s; 0, 0)  \ . \ee 
The   partition function for a CST  field    defined  on \ads{4} is  found to be  \cite{Beccaria:2015vaa} (cf. \rf{2.1})
\be
\label{3.19}
Z_{\text{CST}, \, s}(\ads{4}) = \prod_{k=1}^{s}\Big[
\frac{\det\widehat\Delta_{0}(2-k-k^{2})}{\det\widehat\Delta_{k\perp}(2+k)}
\Big]^{1/2} \ . 
\ee
Using \rf{2.2},\rf{2.4}   we then find for the one-particle partition  function   on thermal \ads{4}
\bea
&&\Z^{+}_{\text{CST}, \, s}(\ads{4}; q) = 
-  \bZ^{-}_{\text{CST}, \, s} (\ads{4}; q^{-1} )
 =    \frac{1}{(1-q)^{3}}\sum_{k=1}^{s}\Big[ (2k+1)\,q^{2}-q^{k+2}
\Big] \ \no \\
&& \qquad \qquad =
\frac{q^{2}\big[s(s+2)-(s+1)^{2}\, q+q^{s+1}\big]}{(1-q)^{4}} \ . \la{3.20}
\eea
This 4d  CST  field   corresponds to   the maximal-depth partially massless (PM)
totally symmetric  spin $s$ field in 
\ads{5} associated with the  following 
 combination of $SO(2,4)$  representations   \cite{Bekaert:2013zya}\footnote{The subscript $r$ 
in $\text{PM}_{s}^{(r)}$ denotes the depth. Here, we  consider only  the maximal case $r=s$.}
\be
\label{3.17}
\text{PM}_{s}^{(s)} = (3;\tfrac{s}{2}, \tfrac{s}{2})-(3+s; 0,0)\ ,
  \ee
 for which \rf{319} is a "shadow" counterpart. 
Then from \rf{2.4} we get 
\be
\label{3.18}
\Z^{+}_{\text{PM}_{s}^{(s)}}(\ads{5}; q) = \bZ^{-}_{\text{PM}_{s}^{(s)}}(\ads{5}; q^{-1} )
 = \frac{(s+1)^{2}q^{3}-q^{s+3}}{(1-q)^{4}}\ .
  \ee
Comparing \rf{3.20} and \rf{3.18}   we conclude that 
\be
\label{3.22}
\bZ_{\text{PM}_{s}^{(s)}}^{-}(\text{AdS}_{5}; q) -\Z_{\text{PM}_{s}^{(s)}}^{+}(\text{AdS}_{5}; q) = 
\bZ_{\text{CST},\, s}^{-}(\text{AdS}_{4}; q) +\Z_{\text{CST}, \, s}^{+}(\text{AdS}_{4}; q) \ , 
\ee
in agreement with \rf{3.1a}. As in CHS  case, one can   
also  verify the validity of \rf{3.1} also  in the  CST  case.
We shall   further discuss the properties of 4d CST   field   partition functions in the next section.

\section{ From  5  to 4 to   3  dimensions 
  }

In this  section we shall consider a special case   of $d=4$ where some relation simplify. 
   We shall  discuss 
further descent to 3 dimensions thus  getting a "triple" of related fields: HS in 5,  CF in 4,  and  $\wCF$  in  3 dimensions. 
 In particular, starting with a massless higher   spin field in \ads{5}   one  gets a conformal higher spin in 4d 
 and then defininig it on \ads{4} can further associate  to it another 
  conformal higher spin  field  in 3d. The latter   turns out  to have 
 a non-local action describing   zero  number of dynamical degrees of freedom, i.e.  
 giving      trivial partition function.


\subsection{\ads{5} $\to $ \ads{4}  } 


Let us  first  consider  the  $d=4$ version of the relation \rf{3.1}   between partition functions of  some 
 higher spin   field in \ads{5}   and the corresponding 4d   conformal field defined on \ads{4}. 
 Starting with a  totally symmetric spin $s$   HS  field  in \ads{5}    corresponding to $SO(2,4)$ 
representation $(\Delta_{5}; \tfrac{s}{2},\tfrac{s}{2})$  
  we may associate to it  (in general, higher-derivative) conformal field in 4d  that may also be represented 
  (when defined  on $\RR^4$ or \ads{4})  as a  collection of  2nd-derivative  fields   with particular  values of masses. 
  Our  proposal for such general relation  is\foot{A similar  "correspondence rule" in 6d 
was discussed   in 
Appendix A of \cite{Beccaria:2015uta}.}
\be
\begin{split}
\label{4.33}
(\Delta_{5}; \tfrac{s}{2},\tfrac{s}{2})_{ \ads{5} }\ \ \   &\longrightarrow  \ \ \ 
Z_{{\rm CF},s} (  \ads{4} ) = \prod_{s'=0}^{s}\prod_{k=0}^{\Delta_{5}-3}
\Big[\det\widehat\Delta_{s'\,\perp}(M^2_{s',k})\Big]^{-1/2}\  , 
\end{split}
\ee
where $M^2_{s',k}= s' -  2  - k (k+1) $  as   in \rf{3.122}.
Special HS fields with gauge invariance  will  require combinations of the above building blocks to take
into account ghost field  contributions.
One  can check that (\ref{4.33}) is consistent with all   special   conformal fields in 4d 
that we  have analysed   directly:
  CHS,  CST  and also  GJMS scalar   fields  (see Appendix \ref{a3}).
  
 Given  \rf{4.33} one can then demonstrate the validity of the relation \rf{3.1} (equivalent to \rf{3.1a} in the absence of gauge invariance) 
   between the partition functions
 in \ads{5} and \ads{4}. 
  For each factor in the r.h.s. of (\ref{4.33}) we find from \rf{2.2}  that $\Delta_{4,k}^{+} = k+2$ 
  and thus applying \rf{2.4}  to both 5d and 4d cases   we get 
\bea
\label{4.34}
&&\Z_{(\Delta_{5}; \tfrac{s}{2},\tfrac{s}{2})}^{+}(\ads{5}; q) = (s+1)^{2}\,\frac{q^{\Delta_{5}}}{(1-q)^{4}}\ , \\
&&\Z_{\text{CF}}^{+}(\ads{4}; q) =\sum_{s'=0}^s \sum_{k=0}^{\Delta_{5}-3}
(2s'+1)\,\frac{q^{k+2}}{(1-q)^{3}} = (s+1)^{2}\,\frac{q^{2}-q^{\Delta_{5}}}{(1-q)^{4}} \ . 
\la{434}
\eea
Then using the expression  \rf{2.6}    for $\bZ^-$ we indeed   verify \rf{3.1}, i.e.   
\be
\label{4.35}
\bZ^{-}_{(\Delta_{5};  \tfrac{s}{2},\tfrac{s}{2})}(\ads{5}; q) 
- \Z^{+}_{(\Delta_{5};  \tfrac{s}{2},\tfrac{s}{2})}(\ads{5}; q)  = 
\bZ^{-}_{\text{CF}}(\ads{4}; q) +\Z^{+}_{\text{CF}}(\ads{4}; q) \ .
\ee
To provide additional support  for the  correspondence rule 
  (\ref{4.33})  
    let us consider  the   CF  partition function   defined on $S^4$ instead of \ads{4}
    which may be viewed as a boundary of global \ads{5}.  In that case we should  get  an analog of \rf{2.7}, i.e. 
      the  relation \rf{1.1} 
    \be
\label{2.77}
\log {Z^{-}_{\rm HS}(\ads{5}) - \log  Z^{+}_{\rm HS}(\ads{5})}= \log Z_{\rm CF}(S^{4})\ .
\ee
One may  check this relation   by  comparing   the  coefficient of the 
 IR  divergent  term on the l.h.s. to the coefficient of the  UV divergent  term on the r.h.s., i.e. to the 4d 
conformal anomaly a-coefficient \ci{Giombi:2013yva,Beccaria:2014jxa}. 
According to Eq.(3.3) of \cite{Beccaria:2014jxa},   we get  for the   coefficient in the  l.h.s.
\be
\label{4.36}
{\text{a}}(\Delta_{5}; \tfrac{s}{2}, \tfrac{s}{2}) = \te 
-\frac{1}{720} (\Delta_{5}-2)^3\, (s+1)^2\,\big[3 \,(\Delta_{5}-2)^{2}-5 s^2-10 s-5\big] \ .
\ee
On the other hand,  each  $\det(-\nabla^{2}_{s}+M^{2})_{s\perp}$
in the product in  (\ref{4.33})  defined on   $S^4$ 
gives the contribution (see    Eq.(3.37) of \cite{Tseytlin:2013jya})
\be
\label{4.37}
{\text{a}}_{s\perp}(M^{2}) = \te  \frac{1}{720}\,(2s+1)\,\big[ 30s^{3}+85s^{2}+10s-58-30(s^{2}-2)M^{2}
-15 M^{4}\big]\ .
\ee
Then for  the  particular combination of  the operators  in  (\ref{4.33})  we get indeed
\be
\label{4.38}
{\text{a}}\big(\Delta_{5}; \tfrac{s}{2}, \tfrac{s}{2}\big) = 
\sum_{k=0}^{\Delta_{5}-3}\sum_{s'=0}^{s}{\text{a}}_{s'\perp}\big(M^{2}_{s',k}\big) \  .
\ee
Let us note that 
the relation (\ref{4.35}), implied   by the correspondence rule (\ref{4.33}),  should have  a group theoretic 
interpretation. 
To see an indication of  this, let us consider the {"non-blind"}
characters $\chi_{4}$ and $\chi_{3}$  of massive representations of $SO(4,2)$ and $SO(3,2)$ respectively
 \cite{Dolan:2005wy}\foot{Here $x$ and $y$ are chemical potentials for  charges corresponding to 
 other Cartan generators in addition  to the  dilatation operator.}
\be
\label{4.39}
\begin{split}
&\chi_{4}(\Delta; j_{1}, j_{2}| q, x, y) = 
\frac{ q^{\Delta}\,f_{\text{SU}(2)}(j_{1}| x)\,f_{\text{SU}(2)}(j_{2}| y)}{
(1-q\,x^{\frac{1}{2}}\,y^{\frac{1}{2}})
(1-q\,x^{\frac{1}{2}}\,y^{-\frac{1}{2}})
(1-q\,x^{-\frac{1}{2}}\,y^{\frac{1}{2}})
(1-q\,x^{-\frac{1}{2}}\,y^{-\frac{1}{2}})
}\ , \\
&\chi_{3}(\Delta; j| q, x) = 
\frac{q^{\Delta}\,f_{\text{SU}(2)}(j| x)}{(1-q)(1-q\,x)(1-q\,x^{-1})}\ ,\qquad\qquad   f_{\text{SU}(2)}(j| x) \equiv 
 \frac{x^{j+\frac{1}{2}}-x^{-j-\frac{1}{2}}}{x^{\frac{1}{2}}-x^{-\frac{1}{2}}}\ .
\end{split}
\ee
 Let us  generalize (\ref{4.34})    and   define 
\be
\label{4.40}
\chi(\ads{5}|q,x) \equiv \chi_{4}(\Delta_{5}; \tfrac{s}{2}, \tfrac{s}{2}| q, x, x)\ , \qquad\qquad 
\chi(\ads{4}|q,x) \equiv \sum_{s'=0}^{s}\sum_{k=0}^{\Delta_{5}-3}
\chi_{3}(k+2; s'| q, x) \ . 
\ee
Let us  also denote by  tilde the  "charge conjugation", i.e.  the   replacement 
$q\to q^{-1}, \  \ x\to x^{-1}$. One can  then check that 
\be
\label{4.41}
\widetilde{\chi}(\ads{5}|q,x)-\chi(\ads{5}|q,x) = \widetilde{\chi}(\ads{4}|q,x)+\chi(\ads{4}|q,x) \ .
\ee
This  reduces  to  (\ref{4.35}) in the "blind" limit $x\to 1$. 
The fact that (\ref{4.41}) holds also  for generic  argument $x$
 suggests  that (\ref{4.35})  has a  group theoretic interpretation in terms of 
  a  map between  representations of the corresponding 5d and 4d   isometry groups.


\subsection{Relation   between  partition functions on \ads{4} and $S^1\times S^3$}

As already  mentioned above, 
  given a conformal field in \ads{d}   we may    make  a Weyl transformation 
to replace \ads{d}    by  half of the Einstein Universe  $R \times S^{d-1}$   and then  
represent the  partition function in  $R \times S^{d-1}$  
  in terms of  the partition function in  \add\  with two   possible choices   of boundary conditions. 
  In the case of thermal quotients   that leads to the relation  \rf{3.2}. 
  
In the special case of \ads{4} 
it was observed in \ci{Breitenlohner:1982bm}  that  the two  choices  (+ and -)  
of the possible  boundary conditions are equivalent, 
i.e.   the corresponding   higher spin representations  
 are equivalent for  $s >0$,    with   spin 0 (scalar)  case  being  an  exception.\foot{More generally, the fact
  that highest weight unitary representation of 
 $\mathfrak{so}(d,2)$ algebra that
admits extension to $\mathfrak{so}(d+1,2)$ conformal algebra has two inequivalent extensions
was demonstrated in \cite{Metsaev:1995jp};
for scalar field these representations
are not equivalent as representations of $\mathfrak{so}(d,2)$ while for spin $s >0$  fields
they  are.} 
This  suggests that  for any  conformal field not containing a scalar component  we should have 
 the equality  between partition functions corresponding to the two alternative boundary conditions
 \bea\la{411}
&& \qquad \qquad  \qquad \qquad \qquad \qquad \Z^{-}_{\text{CF}}(\ads{4}; q)=  \Z_{\text{CF}}^{+}(\ads{4}; q)\ , \\
  && \la{4111} 
  \Z^{-}_{\text{CF}}(\ads{4}; q) =  \widetilde\Z^{-}_{\text{CF}}(\ads{4}; q) + \sigma_4  (q) \ , \ \ \ \qquad 
 \bZ^{-}_{\text{CF}}(\ads{4}; q) = -  \Z^{+}_{\text{CF}}(\ads{4}; q^{-1})\ \ \ \ 
 \ . \eea
 Then  the relation \rf{3.2}   should simplify  in $d=4$   case to 
  \be
   \label{4.1}
 \Z_{{\text{CF}}}(S^{1}\times S^{3}; q) = 2\, \Z_{\text{CF}}^{+}(\ads{4}; q)\ . 
   \ee  
This  identity can be verified  directly 
   for  the  CHS  or CST fields    as follows.
   
 The  CHS  partition function on \ads{4} is a special case of  (\ref{3.12})\footnote{While  here  we 
 have  a scalar field   contribution  at $k=0$  this is not  a conformal massless scalar but 
   a ghost  field needed to guarantee the conformal invariance of  the  spin $s$  CHS field.}
\be
\label{4.2}
Z_{\text{CHS},\, s}(\ads{4}) = \prod_{k=0}^{s-1}\Big[
\frac{\det\widehat\Delta_{k\perp}(k-(s-1)(s+2))}
{\det\widehat\Delta_{s\perp}(s-(k-1)(k+2))}
\Big]^{1/2}\ .
\ee
From \rf{2.2} we see  that the fields   corresponding to 
  terms in the numerator  have  $\Delta_{k}^{+} = s+2$, while
the  terms in the denominator give $\Delta_{k}^{+} = k+2$.
The corresponding one-particle partition function  is then a $d=4$  case of \rf{3.14} 
\be
\label{4.3}
\Z_{\text{CHS},\, s}^{+}(\text{AdS}_{4}; q) = \sum_{k=0}^{s-1}  \te \frac{(2s+1)\,q^{k+2}-(2k+1)\,q^{s+2}}{(1-q)^{3}} 
= \frac{(2s+1)q^{2}-(s+1)^{2}q^{s+2}+s^{2}q^{s+3}}{(1-q)^{4}}\ . 
\ee
Comparing this  to the  CHS  partition function on $S^{1}\times S^{3}$   given in  Eq.~(4.8) of   \cite{Beccaria:2014jxa}
we find that indeed
\be
\label{4.5}
\Z_{\text{CHS},\, s}(S^{1}\times S^{3}; q) = 2\, \Z_{\text{CHS},\, s}^{+}(\ads{4}; q)\ .
\ee
In the case of the  CST   field the partition   function on $S^{1}\times S^{3}$ was found in \cite{Beccaria:2015vaa}. 
Comparing to \rf{3.20}   we conclude   again that 
\be
\label{4.7}
\Z_{\text{CST},\, s}(S^{1}\times S^{3}; q) = 2\, \Z_{\text{CST},\, s}^{+}(\ads{4}; q)\ .
\ee
  The 4d relation (\ref{4.1}) may  be extended  also to the  fermionic CHS fields, see 
 Appendix~(\ref{a4}).

This   relation \rf{4.1}  is not, however,   true    for a conformal scalar   and thus 
also for any conformal theory  with  \ads{4}  partition function 
 containing a conformal scalar  factor.
In particular, it is not true   for the 4-derivative conformal scalar  field as 
follows from the  comparison of (\ref{B.17}) and (\ref{B.19}) in Appendix \ref{a3}. 
Another counter-example is the conformal theory of an antisymmetric  rank 2 tensor  discussed in Appendix~\ref{a5}.\foot{Here the presence of a scalar components    is  apparent also  in the approach developed in \cite{Metsaev:2008ba}.} 


\subsection{Further descent: \ads{4}  $\to$  $  \RR\times S^{2}$}
 Given  a conformal  field  in 4d   related to some higher spin   field in \ads{5} 
   we may  define it  on \ads{4} and then 
     further associate to it another conformal field $\wCF$ at the \ads{4} boundary $ \RR\times S^{2}$. This  gives  a   
     triplet of fields 
\be
\notag
\text{HS on \ads{5}} \longrightarrow \text{CF on \ads{4}} \longrightarrow {{\wCF}} \text{ on }  \RR\times S^{2} \ . 
\ee
Then the  thermal  partition functions of CF and $\wCF$  are related  by   \rf{1.2} or \rf{2.7}, i.e. for the second step we get 
\be
\label{4.8}
 \Z^{-}_{\text{CF}}(\ads{4}; q) - \Z^{+}_{\text{CF}}(\ads{4}; q)= \Z_{\wCF}(S^{1}\times S^{2}; q)  \ . 
\ee
Combining  this   with \rf{411}  we conclude   that  $\wCF$ should have zero  partition function, 
\be \label{4.10}
\Z_{\wCF}(S^{1}\times S^{2}; q)   =0 \ . \ee
Then   the total partition function of the theory  defined on "boundary of the boundary"  is 
 $Z=1$, i.e.      the resulting 3d conformal theory $ \wCF$    should be   "trivial" or  "topological".
We will also define  $\wCF$ on \ads{3}   and then  find, in agreement with \rf{3322},   that 
\be 
 \label{4.100}
\Z^+_{\wCF}(\ads{3}; q)=\Z^-_{\wCF}(\ads{3}; q)   =0 \ . \ee
Before  giving  some explicit examples let us first recall that 
a  totally symmetric  spin $s$  CHS  field  $\p_s$   in   $d$  dimensions    has the action
\be
\label{4.11}
S_{\text{CHS}_{s}} = 
\int d^d x  \,\p_s\, P_s\, \del^{2s +  d-4 }\,   \p_s
=  \int d^d x\,  C_s\, \del^{d-4 }\,   C_s \ ,
\ee
where  $P_{s}$ is a projector 
onto transverse traceless tensors   and 
$C_s \sim   \partial^s \p_s$  is   gauge-invariant field strength (generalized  Maxwell or  Weyl tensor).
The number of  the corresponding dynamical degrees of freedom  is  \ci{Metsaev:2007rw,Tseytlin:2013fca}
\be
\label{4.12}\te 
\nu_{s, d} = \frac{(d-3)(2s+d-2)(2s+d-4)(s+d-4)!}{2\,(d-2)!\,s!} \ . 
\ee
Eq. \rf{4.11}   is  local   for even $d\geq 4$ (where $\del^2=\Box$ enters in positive power)  
but   can be formally  defined also   for  odd $d$. 
The  case of $d=3$     is  special in that the number of dynamical degrees of freedom \rf{4.12}  vanishes, while 
  \rf{4.11}   takes a non-local form 
\be
\label{4.13}
S_{\text{CHS}_{s}} = 
\int d^3 x  \,\p_s\, P_s\, \Box^{s - {1/2}  }\,   \p_s =  \int d^3 x\,  C_s\, \Box^{-1/2\,  }\,   C_s \ . 
\ee
Let us also recall  that  the  CHS action in $d$ dimensions 
may be viewed as  an  induced  one \cite{Tseytlin:2002gz} 
from a free CFT$_d$: 
  if   $\p_s$  is  coupled 
to  a spin-$s$ conserved   current  $J_s$  then the kinetic term   of $\p_s$ is determined by  the 2-point  function 
$\langle J_{s}(x)\,J_{s}(x')\rangle $.\foot{In general, in   3d  there are two possible conformally invariant tensor structures that may appear 
in a  two-point function of a  conserved current $J_{s}$: a non-local parity-even 
 and  a local  parity-odd one  (see, e.g., \cite{Osborn:1993cr,Leigh:2003ez,Giombi:2013yva}).} 
 Insisting on locality one   may consider   a  Chern-Simons type  action for the corresponding 
 3d CHS field  that may be induced  from chiral 3d fermions  
 (see  \ci{Pope:1989vj,Fradkin:1989xt,Nilsson:2015pua}   and \ci{Deser:1981wh,Horne:1988jf,Afshar:2011qw} for $s=2$).
   For a more 
  natural parity-even  case  induced from  a free   3d scalar CFT  
we get in momentum space 
$\langle J_{s}\,J_{s}\rangle =\,\frac{ k_{s}}{\sqrt{p^{2}}}\td P_s(p) \ , 
$
where ${\td P_s(p)}$  is Fourier transform of the transverse traceless projector in \rf{4.11}
(i.e. a symmetrized and traceless product of  $s$ factors of  $(\td P_1)_{\m}^{\n} = \delta_{\m}^{\n}  - {p_{\mu}p^{\nu}\ov p^2} 
$). 
The  corresponding parity even 3d CHS action is then   given  by  (\ref{4.11}).
 
 \iffa
The second term in \ref{4.14} is a conformally invariant contact term, see \cite{Leigh:2003ez}.
Both are "topological'' because the first term has $\nu_{s,3}=0$ from (\ref{4.12}), while the 
second term is a Chern-Simons contribution, see also \cite{Giombi:2013yva}. In our context, it is only the first, parity
even,  term of 5.5a that is relevant. To understand why, one may follow the logic of the last paragraph in Section 1
of  \cite{Tseytlin:2002gz} and generalize in $d$ dimensions.
If we have a free  3d CFT   with complex scalar $\Phi$ and  currents
$J_s$,   then coupling them to $\p_s$    and integrating $\Phi$ out we get
of course    $\p_s \langle J_s\, J_s' \rangle\, \p_s'$   which is of course   equivalent
to above as $J_s = \Phi^{*} \, \partial^s\,  \Phi$,
symbolically.  
\fi 

\subsubsection{Spin 1 } 

Let us  now illustrate \rf{4.10}  turn to some special cases    and start   with  a massless  spin 1 gauge  field in \ads{5} 
that    has 
$\Delta_{5}^{+}=2+s=3$  and is associated with the following combination of $SO(2,4)$ representations
$ 
\rm{MHS}_{1}(\ads{5}) = (3; \tfrac{1}{2}, \tfrac{1}{2})-(4; 0,0)\ .
$ 
The corresponding 4d boundary  field  is the    $s=1$   CHS field, i.e. the standard  Maxwell   theory (cf. \rf{4.11}). 
 Its partition function  when defined on \ads{4}  is a special case of (\ref{3.12})  
\be
\label{4.17}
Z_{\text{CHS}, 1}(\ads{4}) = \Big[\frac{\det\widehat\Delta_{0}(0)}{\det\widehat\Delta_{1\perp}(3)}\Big]^{1/2}\ .
\ee
Here each operator is in turn  associated with a  conformal field   at the $R\times S^2$    boundary (we get 
a  scalar   with $\Delta_{4}^{+} =3$ and a transverse vector with $\Delta_{4}^{+} =2$). 
The corresponding  
combination of $SO(2,3)$ representations $(\Delta_{4}^{+}; j)$ is\footnote{Here $j$ is the 
$SO(3)$ angular momentum. In general, 
 $\text{MHS}_{s}(\ads{4}) = (1+s; s)-(2+s; s-1)$, i.e. it corresponds to a spin $s$  field with gauge invariance with spin $s-1$ parameter. 
The partition function for a massive $SO(2,3)$ representation $(\Delta_{4}; s)$ is   given by  $\Z^{+}_{(\Delta_{4}; s)}(q) = (2\,s+1)\,\frac{q^{\Delta_{4}}}{(1-q)^{3}}$.
}
\be\label{4.19}
\text{CHS}_{1}(\ads{4}) = (2; 1)-(3; 0) = \text{MHS}_{1}(\ads{4})\ .
\ee
Thus   the  field  at the 3d boundary should be the 
 $s=1$ member of the   3d CHS family  \rf{4.13}  with a  non-local  action  (cf.  \cite{Witten:2003ya,Giombi:2013yva})
\be
\label{4.20}
S_{\text{CHS}_{1}} = 
\int d^3 x
\ 
F_{\mu\nu}\, {\Box^{-1/2}}\,F_{\mu\nu} \ . 
\ee
This theory  is effectively  topological, having   no dynamical degrees    of freedom (in agreement with  \rf{4.12}).
One can see this explicitly, e.g.,  by computing the corresponding  partition function in flat 3d space\foot{Here  
 the measure contribution from the decomposition 
$A_{\mu} = A_{\mu\perp}+\partial_{\mu}\phi$ cancels against the kinetic operator contribution.
Note  also that in 3d   one can dualize $F_{\mu\nu}\frac{1}{\Box^{1/2}}\,F_{\mu\nu}$  to a  scalar   with kinetic term $\p {\Box^{3/2}} \p$
but the corresponding partition function is still  1  as the scalar determinant is cancelled by the measure contribution coping from integrating out the auxiliary field $F_{\m\n}$. } 
\be
\label{4.21}
Z_{\text{CHS}_1}(\mathbb R^{3}) =\te  \Big[
\frac{\det\Box}{\det(\partial\Box^{-1/2}\partial)_{1\perp}}
\Big]^{1/2} =  \Big[
\frac{\det\Box}{\det(\Box^{1/2})_{1\perp}}
\Big]^{1/2} \stackrel{\text{3d}}{=} \Big[
\frac{\det\Box}{\det(\Box^{1/2})^{2}}
\Big]^{1/2}=1\ . 
\ee
As  there is no conformal anomaly in odd dimensions  the same should be true 
also for all conformally flat spaces, e.g., \ads{3}
\be
\label{4.222}
Z_{\text{CHS}_1}({\ads{3}})  =1   \ .
\ee 
Defining the 4d  Maxwell field  
on \ads{4} we get  from (\ref{A.8})  
\bea
\label{4.15}
 \bZ^{-}_{\text{CHS}_1}(\ads{4}; q)-\Z^{+}_{\text{CHS}_1}(\ads{4}; q)  &&= 
-\Z^{+}_{\text{CHS}_1}(\ads{4}; q^{-1})-\Z^{+}_{\text{CHS}_1}(\ads{4}; q)\no  \\
&&\te = -\frac{{3}/{q^2}-{1}/{q^3}}{\left(1-{1}/{q}\right)^3}-\frac{3
   q^2-q^3}{(1-q)^3} = -1\ .
\eea
This -1  is precisely what is  removed by the $\sigma_{_{{\text{CHS}_{1}},\, 4}}(q)$ term in (\ref{266})
in agreement with \rf{411}  so that  we get $\Z_{\text{CHS}_1}(S^1 \times S^2;q) =0$
as  a special case of \rf{4.10}.\footnote{Indeed, the partition function  on  $S^1 \times S^{d-1} $
   for the CHS$_{1}$ Maxwell field  
with the action in \rf{4.11}, i.e.  $\int d^d x  \, F^{\m\n} \Box^{d-4 \ov 2} F_{\m\n}$,  was already  computed  
   in  \ci{Beccaria:2014jxa}  with the general  expression being \be \no 
\Z_{\text{CHS}_1}(S^1 \times S^{d-1};q)  \te = 1-\frac{1-d\,q+d\,q^{d-1}-q^{d}}{(1-q)^{d}} \ .
\ee
This   vanishes for $d=3$.} 

 We can  also  explicitly check the 
equality \rf{388} of the $\sigma$-terms.
For the MHS$_{s}(\ads{5})$ theory  the $\sigma_{_{{\text{MHS}_{s}},\,  5}}(q)$
 term for a general spin $s$ may be found in Eq.~(5.5) of  \ci{Beccaria:2014jxa} and for $s=1$
it is  equal to  $1$, i.e. is indeed  the same as  the above  $\sigma_{_{{\text{CHS}_{1}},\, 4}}(q)$.

%

\subsubsection{Spin 2} 

Let us now  consider  the  $s=2$ case, i.e.   start with  the 
MHS theory in \ads{5} describing   massless  rank-2 tensor
 with $\Delta_{5}^{+}=2+s=4$ and  spin 1 gauge invariance parameter, i.e. 
   associated with the following combination of $SO(2,4)$ representations
$\text{MHS}_{2}(\ads{5}) = (4; 1,1)-(5; \tfrac{1}{2}, \tfrac{1}{2}).$ 
The  dual  conformal field  in 4d  is the $s=2$ CHS  theory, {\em i.e.} Weyl  gravity.
Its partition function on \ads{4} is  a special case of \rf{3.19}, i.e. 
\cite{Tseytlin:1984wj,Fradkin:1983zz,Fradkin:1985am}  
\be
\label{4.23}
Z_{\text{CHS}_2}(\ads{4}) = \Big[\frac{\det\widehat\Delta_{0}(-4)\,\det\widehat\Delta_{1\perp}(-3)}
{\det\widehat\Delta_{2\perp}(4)\,\det\widehat\Delta_{2\perp}(2)}\Big]^{1/2}.
\ee
Using  \rf{2.2}  the values of  the   scaling   dimensions $\Delta_{4}^{+}$   corresponding to each   factor  in \rf{4.23} are 
(cf. \rf{2.1}) 
\be
\label{4.24}
\begin{array}{|c|cccc|}
\hline
\text{
}& 
\widehat\Delta_{2\perp}(4) & 
\widehat\Delta_{2\perp}(2) & 
\widehat\Delta_{1\perp}(-3) & 
\widehat\Delta_{0}(-4) \\ \hline
\Delta_{4}^{+} & 2 & 3 & 4 & 4\\
\hline
\end{array}
\ee
This means that the equivalent  combination of  $SO(2,3)$   representations  
 is\footnote{In general, for a partially massless   spin $s$ field we have 
 $\text{PM}_{s}^{(s)}(\ads{4}) = (2; s)-(2+s; 0)$.}  
\be
\label{4.25}
\text{CHS}_2(\ads{4}) =\text{MHS}_{2}(\ads{4})\oplus \text{PM}_{2}^{(2)}(\ads{4})
= [(3; 2)-(4; 1)]\oplus[(2; 2)-(4; 0)]  \ . 
\ee
Indeed,  the Weyl graviton  on \ads{4} is a combination of  Einstein graviton and a partially massless 
spin 2  field  with  scalar gauge invariance
\cite{Fradkin:1981jc,Deser:1983mm,Tseytlin:2013jya}.

 The 3d  conformal theory "induced" by  4d Weyl graviton at the boundary of  \ads{4} 
thus contains two parts. 
From  the Einstein graviton $\text{MHS}_{2}(\ads{4})$ we get a conformally  invariant   
  3d CHS$_{2}$ or  Weyl  theory with  parity-even non-local  linearized  action  (\ref{4.13}), i.e.  
 $\int d^3 x\ C_2 \Box^{-1/2} C_2$  (see also   \cite{Leigh:2003ez}).\foot{The full non-linear 
 action  of Weyl-invariant gravity in 3d   with  quadratic  part  given by $\int d^3 x\ C_2 \Box^{-1/2} C_2$ 
 where $C_2$ is linearized Weyl tensor can be obtained as an induced one 
  corresponding to a conformally coupled scalar in 3d, i.e. 
 as $\log\det (- \nabla^2 + {1\ov 8}R)$. Since there is no Weyl anomaly in 3 dimensions  this   non-local functional of the metric 
 will  be both reparametrization and Weyl invariant.}
\iffa 
 \footnote{In more details, this is the 
conformal gravity theory \cite{Leigh:2003ez} with parity-even non-local 
action (we omit an overall normalization)
\be
\notag
S = \frac{1}{2}\int\frac{d^{3}p}{(2\,\pi)^{3}}\,K_{\mu\nu, \lambda\rho}(p)\,h^{\mu\nu}(-p)\,h^{\lambda\rho}(p), 
\  K_{\mu\nu, \lambda\rho}(p) = \frac{1}{2\,|p|}[
\Pi_{\mu\lambda}\,\Pi_{\nu\rho}+\Pi_{\mu\rho}\,\Pi_{\nu\lambda}-\Pi_{\mu\nu}\,\Pi_{\lambda\rho}
], 
\ee
with $\Pi_{\mu\nu}(p) = p_{\mu}\,p_{\nu}-\delta_{\mu\nu}\,p^{2}$.
}
\fi 
From the partially massless   field $\text{PM}_{2}^{(2)}(\ads{4})$ we get  
$\text{CST}_{2}$ field representing   a  non-unitary 3d  symmetric  tensor $\vp_{\mu\nu}$ with  scalar 
 gauge invariance $\delta \vp_{\mu\nu}=\partial_{\mu}\del_{\nu}\ep$.
  This theory has a non-local  action  $\int d^3 x\ \vp_2 {\rm P}_2  \Box^{1/2}\vp_2$  
 where ${\rm P}_2$ is an appropriate projector ensuring scalar gauge invariance.\foot{Indeed, 
 from (\ref{4.25}) we  find that the canonical dimension of  $\vp_{\mu\nu}$ is $d-\Delta^+ =  3-2=1$.}
 In summary, the 
  conformal field  combination   corresponding   to 
 (\ref{4.25})     on  flat  3d boundary     is\foot{An alternative way of obtaining  (quadratic)  action for this set of 3d fields 
   is  to start  with the  (linearized) Weyl gravity  in  \ads{4} space, specify separate   boundary conditions 
   for the 4d graviton  and  partially massless  mode (in terms of 3d graviton and 3d $\text{CST}_{2}$   field respectively)
   and then   evaluate the 4d action on the solution of the equations of motion.
   This   procedure  may have  non-linear generalization  if   one starts with the full non-linear 4d Weyl gravity action and considers 
   a generic  asymptotically \ads{4}   background.}
\be
\label{4.26}
\wCF(\RR^3 ) =\text{CHS}_2(\RR^3 )\oplus \text{CST}_{2}(\RR^3 )\ .
\ee
Let us now show  that this  system has   zero total  number $\nu$ of dynamical degrees of freedom.
Indeed, the  CHS$_{2}$ field  has $\nu=0$ according to  (\ref{4.12}).
For CST$_{2}$  the flat space partition function  is 
  \be
 \label{4.27}
 Z_{\text{CST}_2}(\mathbb R^{3}) =\te  \Big[
 \frac{\det\Box_{1\perp}(\det\Box_{0})^{2}}
 {(\det\Box^{1/2})_{2\perp}(\det\partial\Box^{1/2}\partial)_{1\perp}}
 \Big]^{1/2} =  \Big[
 \frac{\det\Box_{1\perp}(\det\Box_{0})^{2}}
 {(\det\Box_{2\perp})^{1/2}(\det\Box_{1\perp})^{3/2}}
 \Big]^{1/2}{=} 1.
 \ee
 Here the  numerator is the Jacobian for the change of variables 
$
 \vp_{\mu\nu} =\vp_{\mu\nu}^{\perp}+\partial_{(\mu} V^{\perp}_{\nu)}+ (\partial_{\mu}\partial_{\nu}-
 \frac{1}{3}\delta_{\mu\nu}\partial^{2})\gamma .
 $
 The denominator  is    from the action  $\int d^{3}x\,\vp_2 {\rm P}_2  \Box^{1/2} \vp_2$
 where the scalar component $\g$ drops out due to  gauge invariance. 
 Thus  $\nu ({\text{CST}_2})\big|_{ d=3} = \frac{3}{2}\times 2 + \frac{1}{2}\times 2-(2+2) = 0$. 
 We   have  used that   in three dimensions a contribution $(\det\Box_{s\perp})^{-1/2}$ in the partition function
 is equivalent\footnote{The number of components of a totally symmetric traceless rank-$s$ tensor $\p_s$  in $d$ dimensions is $N_s  = \binom{s+d-1}{s}-\binom{s+d-3}{s-2}$, i.e.   $N_s\big|_{d=3} = 2s+1$.
 The number of components of transverse ($\partial\cdot\phi_{s\perp}=0$) traceless rank $s >0$  tensor  is 
 $N_{s,\perp} = N_{s}-N_{s-1} \stackrel{\text{3d}}{\to } 2$.
 }
  to $(\det\Box_{0})^{-1}$ (i.e. $\nu=2$)  for $s>0$. 
  
 Let us note for completeness   that   (\ref{4.27}) may be generalized to any spin $s>0$  as follows
 \be
 \label{4.30}
 Z_{\text{CST}_s}(\mathbb R^{3}) = \Big[
 \frac{\det\Box_{s-1, \perp}\,\det\Box^{2}_{s-2,\perp}\cdots \det\Box^{s}}
 {\det\Box_{s\perp}^{1/2}\,\det\Box_{s-1,\perp}^{3/2}\cdots\det\Box_{1\perp}^{s-1/2}}
 \Big]^{1/2}    =1\ .
 \ee
 Thus as for a  CHS field,  the total number of  d.o.f. of a 3d  CST   field   is   zero for any $s$: \ \ 
 $
 \nu = 2\, \sum_{n=1}^{s} (n-\ha) -\big(2\,\sum_{n=1}^{s-1}n +s\big) = 0.$
 
 At the level of partition function, the decomposition (\ref{4.25}) implies 
 \be
 \label{4.30-1}\te 
 \Z^{+}_{\text{CHS}_{2}}(\ads{4}; q) = \frac{5\,q^{3}-3\,q^{4}}{(1-q)^{3}}+\frac{5\,q^{2}-q^{4}}{(1-q)^{3}}
 = \frac{q^{2}\,(5+5\,q-4\,q^{2})}{(1-q)^{3}}\ .
 \ee
 We may  compute the partition function of 
 $\wCF = \text{CHS}_{2}\oplus \text{CST}_{2}$ as in \rf{2.7},\rf{266}
 \be
 \label{4.30-2}
 \begin{split}
 \Z_{\wCF}(S^{1}\times S^{2}; q) &=  -\Z^{+}_{\text{CHS}_{2}}(\ads{4}; q^{-1})
 - \Z^{+}_{\text{CHS}_{2}}(\ads{4}; q)+\sigma_{_{\text{CHS}_{2}},\,  4} (q) \\
 &= -{4}(q + {q}^{-1})-7 + \sigma_{_{{\text{CHS}_{2}},\,  4}} (q)   =0\ , 
 \end{split}
 \ee
which is  in agreement with \rf{4.10} (see  also Appendix \ref{a2}). 
 We used that,  as  one can check,   $\sigma_{_{\text{CHS}_{2}}, 4} (q)  = {4}({q} + q^{-1})+7$.\footnote{Here  we can use the
   expression for  $\Z_{\text{CHS}_2}(S^1 \times S^{d-1};q) $
 in Eq.~(5.17) of \ci{Beccaria:2014jxa}   and check that it vanishes for $d=3$.
 }
 Again, we can compare the expression for  $  \sigma_{_{{\text{CHS}_{2}},\,  4}} (q)$ with the $\sigma_{_{{\text{MHS}_{2}}, 5}}(q)$ term 
 in Eq.~(5.5) of  \ci{Beccaria:2014jxa}  for $s=2$ and thus verify the relation \rf{388}.

\section{Concluding remarks}

The "$(\text{AdS/CFT})^{2}=0$" relation (\ref{4.10}) is special to 
 $d=4$ case  because we used (\ref{411}) to obtain it.
  In  $d>4$  case we    expect to find a  more complicated picture.
  For example, suppose we start with an  massless HS   field   in \ads{7}. Then at the boundary we    get   conformal HS field in 6d
   and  can define  it 
  on \ads{6} thus  associating  to it some other  conformal field  CF$_5$ at the boundary of 
  \ads{6}. We can then continue  the descent, i.e. define   CF$_5$   on \ads{5} and associate to it another 
   conformal field  CF$_4$ at the boundary of \ads{5},  etc. One may then look for   some   new  identities 
   between   partition  functions of these fields  in addition to \rf{2.7} and \rf{3.1}. 
   For  example, one can   check    that in the spin 1  case  CF$_5$  appears to be  represented   by a combination 
   of 5d CHS$_1$ field (with non-local action  $ \int d^5 x\,  F_{\m\n} \Box^{1/2}  F_{\m\n} $ 
    as in \rf{4.11})  and an extra field, such that  one  ends up with  CF$_4$    being just   the standard  CHS$_1$ Maxwell field.
   Details of the corresponding  relations  between partition functions remain to be studied. 
   
Given the general relations  \rf{1.2}  and \rf{313} 
between partition functions of particular  higher spin fields    one may apply them 
to  theories containing infinite number of  spins. For example, the Vasiliev-type theory in \ads{d+1} (containing a scalar 
and totally symmetric MHS   spin 1,2,...  fields and   dual to  the 
 singlet sector of free $U(N)$ scalar theory  in 4d)  is  
 naturally  associated  to the CHS theory of all  conformal spins $s=0,1,2,...$   in $d$ dimensions 
 (with linearized action \rf{4.11}).  
 Summing over all spins  the relation \rf{313}   should trivialise. 
Indeed,  for the  MHS theory we find    \ci{Giombi:2014yra}  (spin 0 field  here  has $\Delta^+ = d-2$)
\bea 
&& \la{611} 
\Z^+_{\text{MHS}} (\ads{d+1}; q) =  \sum_{s=0}^\infty \Z^+_{\text{MHS},s} (\ads{d+1}; q)
 = { q^{d-2}  (1 + q) ^2 \ov (1- q) ^{2d-2} }   \ , \qquad \\
&&  \bZ^- _{\text{MHS}} (\ads{d+1}; q) =  (-1)^d \Z^+ _{\text{MHS}} (\ads{d+1}; q^{-1} )=   (-1)^d
\Z^+_{\text{MHS}}(\ads{d+1}; q)   \ . \  \qquad 
\eea
 Thus, e.g., for $d=4$ 
the l.h.s. of \rf{313} vanishes after summing over all spins (the $\sigma$ term in $\Z^-$  in \rf{112} drops out being symmetric 
under $q\to q^{-1}$).
At the same time, the summed CHS partition function  on thermal $\ads{d}$ in \rf{3.14} 
  appears to be divergent  (cf. \ci{Beccaria:2014jxa}). 
   In four dimensions
   $\Z^+_{\text{CHS},s} (\ads{4}; q)$     is given by    \rf{4.3} 
   and the divergence is due to the term $\sim (2s+1)\,q^{2}$ in the numerator that is not suppressed at large $s$.
Nevertheless, applying the analog of the standard  $\zeta$-function regularization
 (i.e. $\sum_{s=1}^{\infty} 1 = \zeta(0) = -\tfrac{1}{2}$, etc.),
we obtain for the regularized   expression of the sum over spins 
\bea
\label{511}   
&& \Z^+_{ \text{CHS}  } (\ads{4}; q)=
\lim_{z\to 0}\sum_{s=1}^{\infty} s^{z}\,\Z^{+}_{\text{CHS},s}(\ads{4}; q)  =
 -{\te \frac{2}{3}}\, \frac{q^2(q^{2}+4q+1)}{(1-q)^{6}}\ , \\
 && \Z^+_{ \text{CHS}  } (\ads{4}; q)=  \Z^+_{ \text{CHS}  } (\ads{4}; q^{-1})\ , \ \ \ \ \ \ \ \
 \bZ^-_{ \text{CHS}  } (\ads{4}; q)= - \Z^+_{ \text{CHS}  } (\ads{4}; q) \ . 
\eea
Thus 
the r.h.s. of \rf{313} also vanishes. The  same  conclusion is   reached  also 
in general even dimension $d>4$  once the   non-trivial  spin 0    contribution is included in the sum in \rf{511}.


\iffa 

 It would  be interesting to explore several directions:
 
 \begin{enumerate}
\item[(i)]   apply this  relation to infinite  sets    of  fields, i.e.  start, e.g.,  with Vasiliev's theory in \addd, associate to it conformal higher spin theory 
 in \add  and then   consider the  corresponding  conformal  theory  in $d-1$ dimensions   that should    represent a collection of  topological    
fields. In fact, each  of these theories in a sense   have trivial partition functions once one sums over all spins \ci{Beccaria:2015vaa} 
and implications of this fact in the resent  context are worth  clarifying further. 

\item[(ii)]  one  may consider both even and odd   values  of $d$   and  both cases  appear to be qualitatively  different;
it would   be  interesting to study,   in particular,    $d=6$  case   (when $\tp$ theory is  4-dimensional) and  also $d=3$ case (when $\tp$ theory is  2-dimensional).

\item[(iii)]  does that imply   interesting relations  for  conformal anomalies in $d$  even dimensions 
  if  viewed  not only from $d+1$ but also from $d-1$  perspective. \footnote{In  the "kinematical"  AdS/CFT one relates the  coefficient of the logarithmic 
  IR divergence of the effective action in \ads{d+1} to  the  coefficient  of the UV   divergence  (or conformal 
  anomaly) in \ads{d}. For  even  $d$  this gives a non trivial  matching, but for odd $d$ both  coefficients are 
   zero.}
  Also, check relations between partition functions for global AdS  when  second  boundary is $S^{d-1}$. 
 
\item[(iv)] can we say something about  asymptotically AdS   cases?   
 May be yes if boundary of asymptotically \ads{d+1}  is asymptotically \ads{d}  itself. 
 Then may be there  are some implications for  relations of curved-space partition functions or their derivatives (correlators). 
 \end{enumerate}

We know that if we use Maxwell   vector  as 4d CFT to define
5d  type-C analog of  MHS Vasiliev theory we get
all sorts  mixed reps in \ads{5}, see the spectrum in table 5 of  \cite{Beccaria:2014zma}.
Then to this spectrum it corresponds also type-C CHS  theory in 4d
for corresponding operators. If Maxwell field is real then we will have  1 scalar (dilaton) coupled
to $F_{\mu\nu} F_{\mu\nu}$  so  it will have  dim 0 and thus CHS action will have its
kin term as
$\phi D^4 \phi  +\dots$. This is same scalar that is present in  $\mc N=4$
conformal  supergravity in 4d ; this scalar replaces non-diagonal    scalar of
usual CHS  theory.
Then if we take   this type-C   CHS theory further to 3d  we
will not have nice   4d property satisfied    because of that scalar present.

we should also add about full AdS/CFT:   we know that 
ZMHS,Vasiliev(ads5) =1  if we sum over all fields; 
then also ZCHS(ads4) =1 is likely 
but  now also    for dual collection of all  fields  in 3d  for *each* mode 
 Z=1  even before summation.

\fi

\acknowledgments
We  thank  M. Grigoriev,   I. Lovrekovic  and   R. Metsaev
 for   useful discussions.  
The work of AAT was  supported   by   the 
ERC Advanced grant No.290456, the STFC Consolidated  grant ST/L00044X/1  and 
by  the  Russian Science Foundation grant 14-42-00047  associated with Lebedev Institute.


\appendix

\section{Reconstructing $\Z^{+}_{\text{CF}}(\ads{d})$ from $\Z^{+}_{\text{HS}}(\ads{d+1})$ }
\label{a1}

Here   we shall  reverse   the logic: assume that the relation \rf{3.1a}  is true  and use it to determine 
the partition function $\Z^{+}_{\text{CF}}(\ads{d})$ from the knowledge  of $\Z^{+}_{\text{HS}}(\ads{d+1})$
just by doing algebraic  manipulations.

 These    partition functions   have the following general  form 
\be
\label{A.1}
\Z^{+}_{\text{HS}}(\ads{d+1}; q) = \frac{P(q)\,q^{\frac{d}{2}}}{(1-q)^{d}}, \qquad\qquad 
\Z^{+}_{\text{CF}}(\ads{d}; q) = \frac{F(q)\,q^{\frac{d-1}{2}}}{(1-q)^{d-1}},
\ee
where $P(q)$ and $F(q)$ are finite sums of non-negative powers of 
$q$ (in \ads{d} we have $\Delta^{+}_d \ge \ha (d-1)$, cf. \rf{2.4}).
 Eq. (\ref{3.1a}) or \rf{3.8}   implies that 
\be
\label{A.2}
F(q)+F(q^{-1}) = \frac{\sqrt{q}}{1-q}\Big[P(q^{-1})-P(q)\Big] \ .
\ee
The r.h.s. of (\ref{A.2}) may be expanded in a Laurent series around $q=0$  and 
comparing with the l.h.s., we  may then  determine $F(q)$.

Let us  consider  some examples. 
Let us  start with  the  conformal scalar in $\ads{4}$  which corresponds  to a massive scalar 
in \ads{5} with $\Delta^{+}_{5}=3$, i.e. (cf. \rf{3.3})  
\be
\label{A.3}
\Z^{+}_{0}(\ads{5}; q) = \frac{q^{3}}{(1-q)^{4}}\  ,  \quad \ \quad \qquad  P(q) = q\ .
\ee
Then  (\ref{A.2}) gives
\be
\label{A.4}
F(q)+F(q^{-1}) = \te \frac{q+1}{\sqrt q} = q^{-1/2}+q^{1/2}\quad \longrightarrow \quad F(q) = q^{1/2}\ ,
\ee
and therefore,  in agreement with \rf{3.6}, 
$
\Z^{+}_{\text{cs}}(\ads{4}; q) = \frac{q^{2}}{(1-q)^{3}}.$

Another example is the spin 1  field in \ads{5} corresponding to  spin 1 
 CHS field (i.e.  Maxwell field) in \ads{4}. 
 From (\ref{3.11}) we have 
\be
\label{A.7}
\Z^{+}_{\text{MHS}_1}(\ads{5}; q) = \frac{4q^{3}-q^{4}}{(1-q)^{4}}\ , \quad \qquad  \qquad P(q) = 
4\,q-q^{2}.
\ee
Then, (\ref{A.2}) gives
\be
\label{A.8}
F(q)+F(q^{-1}) = -q^{3/2}-q^{-3/2}+3\,q^{1/2}+3\,q^{-1/2}\quad \longrightarrow \quad F(q) = 3\,q^{1/2}-q^{3/2},
\ee
and therefore  we get,  in agreement with (\ref{3.14}), 
$
\Z^{+}_{\text{CHS}_1}(\ads{4}; q) = \frac{3\,q^{2}-q^{3}}{(1-q)^{3}} .$

Our third  example is a non-unitary  CFT   represented  by 
a   vector $V_{\mu}$   in 6d  with 2nd-derivative kinetic term. 
This is a special 
$s=1$  case of  CST   family of conformal fields  described by 
rank-$s$ symmetric tensors $\varphi_{\mu_{1}\dots \mu_{s}}$ which  in $d=6$   have   no gauge invariance.
As discussed in \cite{Beccaria:2015uta}, this CF is induced by a higher spin field in \ads{7}
transforming in the $(\Delta; h_{1}, h_{2}, h_{3})=(4; 1,0,0)$ representation of $SO(2,6)$.  Taking into account
that $\dim[1,0,0]=6$,    we get (cf. \rf{2.4})
\be
\label{A.10}
\Z^{+}_\text{HS}(\ads{7}; q) = \frac{6\,q^{4}}{(1-q)^{6}}\ ,  \quad \qquad \qquad P(q) = q.
\ee
Then from  (\ref{A.2})  we have again 
$F(q) = q^{1/2}$ as in (\ref{A.4}) and thus  we  predict that 
\be
\label{A.11}
\Z^{+}_{V}(\ads{6}; q) = \frac{6\,q^{3}}{(1-q)^{5}}\ .
\ee
This   expression   follows  indeed   from  the general 
expression   for  the $V_\m$ partition function on $S^{6}$ or  on 
 \ads{6}
given in  Eq.~(\ref{A.4}) of \cite{Beccaria:2015uta} 
\be
\label{A.12}
Z_{V}(\ads{6}) = 
\Big[{\det\widehat\Delta_{1\perp}(7)\,\det\widehat\Delta_{0}(6)}\Big]^{-1/2}\ .
\ee
Applying (\ref{2.2})  to  the operators   here  we find that  we have  the same  $\Delta^{+}_{6}=3$ and
$\Delta^{-}_{6}=2$ for both   factors. However, the number of degrees of freedom  of a transverse vector in 6d is $6-1=5$
while the scalar contributes only one. Thus  the  numerator of $\Z^{+}_{V}(\ads{6}; q)$ should  
   be $(5+1)\,q^{3}=6\,q^{3}$, in agreement    with \rf{A.11}.

\iffa 
We conclude that   given  \rf{3.1a}  the expression   for 
 $\Z^{+}_{\text{HS}}(\ads{d+1}; q)$    formally  determines (just by algebraic  manipulations) 
$\Z^{+}_{\text{CF}}(\ads{d}; q)$.

If we can prove that (\ref{D.1}) holds for the theory under consideration, this also implies
\be
\label{A.14}
\Z^{+}_{\text{HS}}(\ads{d+1}; q) \qquad \text{determines} \qquad \Z_{\text{CF}}(S^{1}\times S^{d-1}; q).
\ee
Actually, we already know that  (\ref{A.14}) is true in the following sense. According to (\ref{2.7}), one
can evaluate the difference
$\Z^{-}_{\rm HS}(\ads{d+1}; q)-\Z^{+}_{\rm HS}(\ads{d+1}; q)$. Then, the $\sigma$ term is identified by 
removing all negative powers of $q$ and taking into account that $\sigma(q)$ is symmetric under $q\to q^{-1}$.
Thus, (\ref{A.14}) does not allow to do more computations than what we already know. However, if (\ref{D.1})
can be shown to hold in general, then (\ref{A.14}) bypasses the analysis of $\sigma(q)$ and is somewhat more direct.
To explain more concretely this point, consider again the example of spin 1 in (\ref{A.7}). We have 
\be
\begin{split}
\label{A.15}
\Z^{-}_{\text{MHS}, 1}(\ads{5}; q) & -\Z^{+}_{\text{MHS}, 1}(\ads{5}; q) = 
\Z^{+}_{\text{MHS}, 1}(\ads{5}; q^{-1})-\Z^{+}_{\text{MHS}, 1}(\ads{5}; q)  \\
&= \frac{-1+3\,q+3\,q^{2}-q^{3}}{(1-q)^{3}} = -1+6\,q^{2}+16\,q^{3}+30\,q^{4}+\cdots,
\end{split}
\ee
leading to $\sigma(q) = 1$ and, using (\ref{2.7}),
\be
\label{A.16}
\Z_{\rm Maxwell}(S^{1}\times S^{3}; q) = \frac{-1+3\,q+3\,q^{2}-q^{3}}{(1-q)^{3}}+1 = 
\frac{2\,(3\,q^{2}-q^{3})}{(1-q)^{3}},
\ee
in agreement with (\ref{A.9}) and (\ref{D.1}).
\fi

\def \wh  {\widehat}
\def \wDelta  {\wh \Delta} 


\section{$\s$-term relation in Eq. \rf{388} }
\label{a2}

Let us now use  the general  general  structure \rf{A.1},\rf{A.2}    of the  higher  spin  partition functions in \ads{d+1}  and 
the corresponding conformal field  partition functions 
in \ads{d} to   justify  the equality  in \rf{388}.

According to \rf{266},\rf{2.7}   we have   for the  partition  function of CF  on 
 $S^{1}\times S^{d}$ and   the   partition of     another  conformal field $\wCF$  (the one associated to  CF  on \ads{d}) 
 on  $S^{1}\times S^{d-1}$
\bea
\label{X.3}
&&\Z_{\text{CF}}(S^{1}\times S^{d}; q) = 
\widetilde{\Z}^{-}_{\text{HS}}(\ads{d+1}; q) - \Z^{+}_{\text{HS}}(\ads{d+1}; q) 
+  \sigma_{_{{\rm HS}, \, {d+1}}}(q)\ , \\
&&\Z_{\wCF}(S^{1}\times S^{d-1}; q) = 
\widetilde{\Z}^{-}_{\text{CF}}(\ads{d}; q) - \Z^{+}_{\text{CF}}(\ads{d}; q) 
+ \sigma_{_{{\rm CF}, \, {d}}}(q)   \ . \la{X4} 
\eea
Using $\widetilde{\Z}^{-}(\ads{n}; q) = (-1)^{n+1}\,\Z^{+}(\ads{n}; q^{-1})$, and also (\ref{A.2}), we find
\bea
\label{X.4}
&&\Z_{\text{CF}}(S^{1}\times S^{d}; q) = 
\frac{q^{\frac{d-1}{2}}\,[F(q^{-1})+F(q)]}{(1-q)^{d-1}} 
+  \sigma_{_{{\rm HS}, \, {d+1}}}(q)  \ , \\
&&\Z_{\wCF}(S^{1}\times S^{d-1}; q) = 
\frac{q^{\frac{d-1}{2}}\,[F(q^{-1})-F(q)]}{(1-q)^{d-1}} 
+  \sigma_{_{{\rm CF}, \, {d}}}(q)  \ . \la{X44} 
\eea  
The role of the $\sigma$-terms is to remove the  negative powers of $q$ in the expansion around $q=0$  of  the r.h.s. of \rf{X.4},\rf{X44} 
as such terms cannot be present on the l.h.s.  that  can be computed using operator counting method and thus should 
 have only positive powers  of $q$. 
Such terms 
come  only  from the $F(q^{-1})$ term that is the same in the two lines of (\ref{X.4}). This then implies that 
\be
\label{X.5}
 \sigma_{_{{\rm HS}, \, {d+1}}}(q)=  \sigma_{_{{\rm CF}, \, {d}}}(q)\ . 
\ee
To   give an   example, let us consider the CHS$_{2}$ field 
 in 4d  for which  from \rf{4.30-1} we have 
\be
\label{X.6}
\Z^{+}_{\text{CHS},2}(\ads{4}; q)\te  = \frac{q^{2}\,(5+5q-4q^{2})}{(1-q)^{3}}\ \ \  \longrightarrow \  \ \
F(q) = 5\,q^{1/2}+5\,q^{3/2}-4\,q^{5/2}.
\ee
Then  (\ref{X.4}) contains 
\be
\label{X.7}
\frac{q^{3/2}\,F(q^{-1})}{(1-q)^{3}}\te  = \frac{-4+5\,q+5\,q^{2}}{q\,(1-q)^{3}} = -{4}{q}^{-1} -7-4\,q
+5\,q^{2}+\cdots,
\ee
leading to the same result as in \ci{Beccaria:2014jxa}
\be
\label{X.8}  \sigma_{_{{\rm MHS}_2}, \, {5}}(q) = {4}{q}^{-1} +7+4\,q \ ,
\ee
with the same expression  also for $\sigma_{_{{\rm CHS}_2}, \, {4}}(q)$.

\section{Higher derivative conformal scalar  fields}
\label{a3}

Here we will illustrate the relations \rf{3.1},\rf{3.2},\rf{3.1a}  on the example 
of  Weyl-covariant  scalar theory  with     kinetic operator 
$\wh \Delta_{(2r)} = -(\nabla^2)^{r}+\dots$  where dots stand for curvature  dependent terms. 

The  GJMS  
operators $\wh \Delta_{(2r)} $   naturally exist in a 
 technical sense  for $r\le d/2$ (see \ci{Juhl:2011aa}  and refs. there).This means   that  their definition in generic dimension $d$ involves terms whose coefficients have  
poles in  $d$ when $r>d/2$. For instance, for $r=3$, there are tensor structures with coefficient
$1\ov d-4$ forbidding a naive  extension to  4d case. These  obstructions are proportional to Bach tensor  and  vanish  for the 
Einstein spaces with $R_{mn} = \frac{R}{d}\,g_{mn}$.
In this case  it is 
possible to construct  generalized Gover-GJMS operators (defined 
beyond critical order)  but  the resulting expressions are non-natural  in technical sense \cite{Gover:2005mn}. For all
orders (below, at, and beyond critical order) the  Gover-GJMS operators  factorise as   in (\ref{B.14}) below.
In the  conformally flat spaces (that need not be Einstein in general), there is no obstruction  in going beyond the critical order $r=d/2$. 
%

\def \GJMS  {{\rm GJMS}}

\subsection{Partition function on $S^{1}\times S^{3}$
}

\def \OO  {{\cal O}}

Let us first   consider the  general case of the 
 space $S^{q}\times S^{p}$  
 is conformally  flat if defined 
 with indefinite $(p,q)$ signature  metric (here the spheres   have  unit radius   and $d=p+q$). Then  
 one    can show that for $r= 2N$ \cite{Juhl:2009aa} 
\bea
&&\label{B.1}
\wDelta_{(4N)} = \prod_{k=1}^{N}\left[
(\OO_p-\OO_q)^{2}-2(2k-1)^{2}(\OO_p+\OO_q)+(2k-1)^{4}
\right]\ ,   \\
&&\wDelta_{(4N+2)} = (\OO_p- \OO_q)\,\prod_{k=1}^{N}\left[
(\OO_p-\OO_q)^{2}-2\,(2k)^{2}(\OO_p+\OO_q)+(2k)^{4}
\right] \ \label{B.111}
\eea
where $\OO_p \equiv  -\nabla^2_{S^{p}}+\fo ({p-1}{})^{2} $.
For example, in  the special  case of $\Delta_{(4)}$ in $d=4$ we have  for a general  curved  background 
\ci{Fradkin:1981jc,Paneitz:1983}
\be
\label{B.6}
\wDelta_{(4)} \te= -(\nabla^{2})^{2}+2\,(R^{mn}-\frac{1}{3}g^{mn}\,R)\,\nabla_{m}\nabla_{n}\ . 
\ee
Then  for 
  $S^{2}\times S'^{2}$   with $(++--)  $  signature we get\foot{Here $
  R_{mn}=\pm g_{mn} = g_{mn}^{(0)}$, where $g_{mn}^{(0)}$ is the metric 
of a standard 2-sphere and the sign depends on whether  $(mn)$ are in the first or second
sphere. Thus $ \nabla^{2} =  -\nabla^2_{S^{2}}+\nabla^2_{S^{2}}$ and 
$R^{mn}\nabla_{m}\nabla_{n} =\nabla^2_{S^{2}}+\nabla^2_{S^{2}}$. 
} 
 \be
\label{B.3}
\wDelta_{(4)} =( -\nabla^2_{S^{2}}+\nabla^2_{S'^{2}})^2   + 2 (\nabla^2_{S^{2}}+\nabla^2_{S'^{2}}) = 
(\OO_2- \OO'_2)^{2}-2(\OO_2  + \OO'_2)+1\  ,\ee
in agreement with \rf{B.1} where $    
\OO_2 = -\nabla^2_{S^{2}}+\frac{1}{4}. $

For conformally flat but  not Einstein   space   $S^1 \times S^3$  with  $(-+++)  $  signature 
we have  for  the   spectra   of $\OO_p$ in \rf{B.1}  
\be
\label{B.8}
\OO_1 \to  w  \ ,\qquad \qquad   \OO_3 \to  n(n+2)+1 = (n+1)^{2} \ , \ \  n=0,1,2, ...
\ee
For $r=2N$ the factorisation of $\Delta_{(4N)}$   in \rf{B.1}  leads to the  energy   eigenvalues  $w$  represented by 
\iffa 
 the spectrum
\be
\label{B.9}
w = n+2k, \ \ n+2-2k, \qquad k = 1, \dots, N=r/2.
\ee 
This are the sequences
\be
\label{B.10}
\begin{split}
w = n, n+2, n+4, \dots, n+r, \\
w = n-2, n-4, \dots, n+2-r
\end{split}
\ee
Joining them, we conclude that the 
spectrum is 
\fi
\be
\label{B.11}
w_n = n+2-r+2k\ ,\qquad  \qquad k=0, \dots, r-1 \ , 
\ee
so that the  final   one-particle partition function for a GJMS scalar field  is 
\be
\label{B.12}
\Z_{\GJMS_r}(S^{1}\times S^{3}; q) = \sum_{n=0}^{\infty}\sum_{k=0}^{r-1}(n+1)^{2}\,q^{n+2-r+2k} = \frac{q^{2-r}-q^{2+r}}{(1-q)^{4}}
\  .
\ee
This is  the same as  the  partition function  that  counts   descendants of 
a  conformal  scalar operator in flat space     modulo its equation  of motion.
In  general dimension $d$, for a GJMS scalar $\p$ with canonical dimension $\frac{1}{2}(d-2r)$
and equations of motion $\partial^{2r}\p=0$  of  complementary  dimension $\frac{1}{2}(d+2r)$   we get
\be
\label{B.19}
\Z_{\text{GJMS}_r}(S^{1}\times S^{d-1}; q) = \frac{q^{\frac{d}{2}-r}-q^{\frac{d}{2}+r}}{(1-q)^{d}}\ .
\ee

\def \na {\nabla}

\subsection{Partition function on   \ads{d} }
\label{aa4}

The  discussion in section \ref{subsec:confsc}
may be generalized by considering 
a  massive scalar field  in \ads{d+1} with $\Delta^{\pm}_{d+1} = \frac{1}{2}(d\pm 2r)$ \  (with $r=2, 3, \dots$)
 for  which the  associated $d$-dimensional boundary conformal  field  is the  higher derivative   conformal scalar 
 with canonical dimension $\Delta^{-}_{d+1} = \frac{1}{2}(d- 2r)$ and the kinetic operator ${\wDelta_{(2r)}}$. 
 Here (\ref{3.3}) is replaced by 
\be
\label{B.13}
\Z^{\pm}_{0,r}(\ads{d+1}; q) = \frac{q^{\frac{1}{2}(d\pm 2r)}}{(1-q)^{d}} \ .
\ee
On a generic $d$-dimensional Einstein space  ${\wDelta_{(2r)}}$   factorizes as follows \ci{Gover:2005mn} 
\be
\label{B.14} {\wDelta_{(2r)}}= 
\prod_{k=1}^{r} \te \Big(-\na^{2}+\frac{(\frac{d}{2}-k)(\frac{d}{2}+k-1)}{d(d-1)}\, R\Big)\ . 
\ee
For  \ads{d}   case we   have  $R = -d(d-1)$
and using (\ref{2.2}) then find  that  for each massive Laplacian factor  in \rf{B.14}  the   associated dimensions  are 
\be
\label{B.15}
\Delta_{d, k}^{+} =\te  \frac{d+2k}{2}-1, \qquad
\Delta_{d, k}^{-} = \frac{d-2k}{2}, \qquad k=1, \dots, r.
\ee
Hence, (\ref{3.6}) is generalized to (there is no gauge invariance so  that $\s_d=0$ and $\bZ^-=\Z^-$)  
\bea
\label{B.17}
&&\Z^{+}_{\GJMS_r}(\ads{d}; q) = \sum_{k=1}^{r} \frac{q^{\Delta_{d, k}^{+}}}{(1-q)^{d-1}}= \frac{q^{d/2}(1-q^{r})}{(1-q)^{d}}\ , \\
&&\Z^{-}_{\GJMS_r}(\ads{d}; q)  = (-1)^{d+1}\, \Z^{+}_{\GJMS_r}(\ads{d}; q^{-1}) \ . \label{B.177}
\eea
Comparing \rf{B.13}   to \rf{B.17},\rf{B.177} 
one  checks again   the relation \rf{3.1},\rf{3.1a}, \rf{3.1} 
\be
\label{B.18}
\Z^{-}_{0,r}(\ads{d+1}; q) - \Z^{+}_{0,r}(\ads{d+1}; q)  = 
\Z^{-}_{{\GJMS}_r}(\ads{d}; q) +\Z^{+}_{{\GJMS}_r}(\ads{d}; q)\ .
\ee
We can also   demonstrate  the relation \rf{3.2}   by  using \rf{B.19} and \rf{B.17},\rf{B.177}  to check that 
 \be
\label{B.20}
\Z_{\text{GJMS}_r}(S^{1}\times S^{d-1}; q) = \Z^{-}_{\text{GJMS}_r}(\ads{d}; q) +\Z^{+}_{\text{GJMS}_r}(\ads{d}; q)\ .
\ee

\def \s  {{\rm s}} 

\section{Fermionic  conformal higher spin  fields}
\label{a4}

The  discussion of partition function relations   for the 
bosonic CHS fields may be extended to the 4d fermionic ones  (see \cite{Tseytlin:2013jya}).
These are boundary conterparts for the 
  massless  spin $s$    fermionic  higher spin   fields in  \ads{5}   to the 
$SO(2,4)$ representation
\be
\label{C.1}
\text{MHS}_{\s} = \left(\s+\tfrac{5}{2}; \tfrac{\s}{2}, \tfrac{\s+1}{2}\right)
+\left(\s+\tfrac{5}{2}; \tfrac{\s+1}{2}, \tfrac{\s}{2}\right), \ \ \ \ \qquad \s\equiv  s- \ha =0, 1, 2, \dots \ . 
\ee
Here $\s=0$  corresponds  to spin $\tfrac{1}{2}$ fermion,  $\s=1$  to conformal gravitino,  etc. 
Recalling that for the massless   $SO(2,4)$ representation 
$(2+j_{1}+j_{2};  j_{1}, j_{2})$ we  have \footnote{The gauge subtraction in (\ref{C.2}) is present only for $\s>0$ but since it happens to 
vanish for $\s=0$ this  formula is general.}
\be
\label{C.2}
\Z^{+}_{(2+j_{1}+j_{2}; j_{1}, j_{2})}(\ads{5}; q) = \frac{q^{2+j_{1}+j_{2}}}{(1-q)^{4}}\Big[
(2j_{1}+1)(2j_{2}+1)-4\,q\,j_{1}\,j_{2}\Big]\ ,
\ee
we obtain (see Eq.~(2.17) in  \cite{Beccaria:2014xda})
\be
\label{C.3}
\Z^{+}_{\text{MHS}_{\s}}(\ads{5}; q) = \frac{2\,(\s+1)(\s+2)\,q^{\frac{5}{2}+\s}-2 \s (\s+1)\,q^{\frac{7}{2}+\s}}{(1-q)^{4}}.
\ee
Applying the "reconstruction" algorithm in Appendix~(\ref{a1}) or   using the  explicit factorized  form 
of the  fermionic CHS  partition function on $S^{4}$ \cite{Tseytlin:2013jya}   and thus also on \ads{4}   one may   find  that 
\be
\label{C.4}
\Z^{+}_{\text{CSH}_{\s}}(\ads{4}; q) = 2\,
\frac{(\s+1)\,(q^{\frac{3}{2}}+q^{\frac{5}{2}})-(\s+1)(\s+2)\,q^{\frac{5}{2}+\s}+\s(\s+1)\,q^{\frac{7}{2}+\s}}
{(1-q)^{4}}.
\ee
One then concludes that 
the relation (\ref{3.1a}) is  again satisfied
\be
\label{C.5}
\bZ^{-}_{\text{MHS}_{\s}}(\ads{5}; q) - \Z^{+}_{\text{MHS}_{\s}}(\ads{5}; q)  = 
\bZ^{-}_{\text{CHS}_{\s}}(\ads{4}; q) +\Z^{+}_{\text{CHS}_{\s}}(\ads{4}; q)\  .
\ee
In addition, the partition function $\Z_{\text{CHS}_{\s}}(S^{1}\times S^{3}; q)$  was found in  
 \cite{Beccaria:2014xda}  (see Eq.~(2.26) there). Comparing it with (\ref{C.4}),  we  conclude  that  the relation 
 (\ref{4.1}) holds  also  for  the
fermionic $\text{CSH}_{\s}$ family, i.e. 
\be
\label{C.6}
\Z_{\text{CHS}_{\s}}(S^{1}\times S^{3}; q) = 2\, \Z^{+}_{\text{CHS}_{\s}}(\ads{4}; q)\  .
\ee

\iffa 
\section{A sum rule}
\label{app:sumrule}

Suppose that we are dealing with a 4d  conformal theory such that (\ref{4.1}) holds.
From (\ref{4.1}) and (\ref{3.2}), we obtain 
\be
\label{D.1}
\Z^{+}(\text{AdS}_{4}; q)-\Z^{-}(\text{AdS}_{4}; q) = \sigma(q),
\ee
and, using (\ref{2.6})
\be
\label{D.2}
\Z^{+}(\text{AdS}_{4}; q)+\Z^{+}(\text{AdS}_{4}; q^{-1}) = \sigma(q).
\ee
Remembering that $\sigma(q)$ is a polynomial in $q+q^{-1}$, we see that the l.h.s. of  (\ref{D.2}) is regular
around $q=1$. A generic term in $\Z^{+}(\text{AdS}_{4}; q)$ is $q^{\Delta^{+}}/(1-q)^{3}$ and since 
\be
\label{D.3}
\frac{q^{\Delta^{+}}}{(1-q)^{3}}+\frac{q^{-\Delta^{+}}}{(1-q^{-1})^{3}} = (2\Delta^{+}-3)\,\Big(
-\frac{1}{(1-q)^{2}}+\frac{1}{1-q}\Big)+\text{regular},
\ee
we conclude that (\ref{D.2}) implies the sum rule
\be
\label{D.4}
\sum_{\rm fields}(-1)^{\rm gh}(2s+1)\,(\Delta^{+}-\tfrac{3}{2}) = 0,
\ee
where $(-1)^{\rm gh}$ is $1/-1$ for physical/ghost fields. Just to test it, for CHS fields
we obtain from  (\ref{3.13}) that  (\ref{D.4}) reads
\be
\label{D.5}
\sum_{k=0}^{s-1}\Big[(2s+1)\,(k+2-\tfrac{3}{2})-(2k+1)\,(s+2-\tfrac{3}{2})\Big] = 0,
\ee
that is indeed trivially true because each term in the sum vanishes separately. The analogue of (\ref{D.4}) in $d>4$ is readily seen to be false. This is consistent with the fact that  the second equation in 
\rf{3.20}
is  true only in  $d=4$.
For CST fields, from (\ref{3.19}) we obtain (\ref{D.4}) in the form 
\be
\label{D.6}
\sum_{k=1}^{s}\Big[(2k+1)(2-\tfrac{3}{2})-(k+2-\tfrac{3}{2})\Big]=0,
\ee
where again all terms are separately zero.
\fi

\section{Conformal antisymmetric tensor fields in 4d
}
\label{a5}

The Weyl-covariant  Lagrangian for  the   conformal 
 antisymmetric tensor field $T_{\m\n}$ on a  generic curved  4d 
background is \cite{Fradkin:1985am}
\be
\label{E.1}\te 
\mathscr L = (\na^{\mu}T_{\mu\nu})^{2}-\frac{1}{4}\,(\na_{\mu}T_{\rho\sigma})^{2}-R_{\mu\nu}
T^{\mu\lambda}T^{\nu}_{\lambda}+\frac{1}{8}R\,T^{2}_{\mu\nu}
+\frac{1}{2}\,R_{\mu\alpha\nu\beta}T^{\mu\nu}T^{\alpha\beta}.
\ee 
  This   conformal field  in flat 4d space corresponds in  \ads{5}  to  a 
  massive spin 1 theory with representation content 
${\rm HS} = (3; 1,0)\oplus(3; 0,1)$  and  no gauge invariance.
 The 
Lagrangian (\ref{E.1}) restricted to  \ads{4}  background 
 gives  the  kinetic operator that     factorizes into  vector   operators 
  as discussed in   \cite{Fradkin:1983zz}. 
The thermal partition function on $S^{1}\times S^{3}$ may be found in eq. (B.26) of \cite{Beccaria:2014jxa}.
As a result, 
\be
\label{E.2}\te
\Z_{\rm HS}^{+}(\ads{5}; q) = \frac{6\,q^{3}}{(1-q)^{4}}\ , \qquad
\Z_{\rm T}^{+}(\ads{4}; q) = \frac{6\,q^{2}}{(1-q)^{3}}\ , \qquad
\Z_{\rm T}(S^{1}\times S^{3}; q) = \frac{6\,q-6\,q^{3}}{(1-q)^{4}}\ .
\ee
One  finds  then that \rf{3.1a} is satisfied 
\be
\label{E.3}
\bZ^{-}_{\rm HS}(\ads{5}; q) - \Z^{+}_{\rm HS}(\ads{5}; q)  = 
\bZ^{-}_{\text{T}}(\ads{4}; q) +\Z^{+}_{\text{T}}(\ads{4}; q)\ ,
\ee
but there is no analogue of (\ref{4.1}). 

Let us elaborate on the derivation 
of $\Z^{+}_{\rm T}(\ads{4}; q)$ in \rf{E.2}. 
The antisymmetric tensor 
partition function on $S^{4}$ is   \cite{Fradkin:1983zz}
\be
\label{E.4}
Z_{\rm T}(S^{4}) =
\left[\det\widehat\Delta_{(1,0)}(4)\det\widehat\Delta_{(0,1)}(4)
\right]^{-1/2} \ ,
\ee
where 
the 2nd order  operator $\widehat\Delta_{(j_{1},j_{2})}(M^{2})$   (cf. \rf{2.1}) 
 acts  on a field in an irreducible $SO(1, 3)$ representation $(j_{1},j_{2})$ (see, e.g., 
  \cite{Christensen:1978md, Tseytlin:2013jya}).
  Similar  partition function
 is found  on \ads{4} where the mass term is related the  corresponding  conformal 
  dimension $\Delta^{\pm}_{4}$ by the 
 following generalization of (\ref{2.2}) \footnote{The explicit dependence on $j_{1,2}$ 
 through the Casimir of $SO(4)$ comes from the 
 particular  definition of the operator, see Eq. (3.5) of \cite{Tseytlin:2013jya}.}
 \be
 \label{E.5}
 \Delta_4^{\pm}(\Delta_{4}^{\pm}-3)-j_{1}\,(j_{1}+1)-j_{2}(j_{2}+1) = -M^{2} \ .
 \ee
 For $M^{2}=4$ and $(j_{1},j_{2})=(0,1)$ or $(1,0)$ as  in (\ref{E.4})
  this gives $\Delta^{+}_{4}=2$  and $\Delta^{-}_{4}=1$. Therefore, 
   $\Z^{+}_{\rm T}(\ads{4}; q) = \frac{6q^{2}}{(1-q)^{3}}$, 
   in  agreement with  (\ref{E.2})  (the factor
 $6$ is the dimension of $(1,0)\oplus(0,1)$ representation). 
 
 Note that the partition function in (\ref{E.4}) leads   to the correct
  value of the   conformal  a-anomaly coefficient  for $T_{\m\n}$. 
 Using  Eqs.~(3.34)-(3.35) of \cite{Tseytlin:2013jya}, the contribution to the a-anomaly from 
 $\widehat\Delta_{(j_{1},j_{2})}(M^{2})$ is  found to be 
 \be
 \label{E.6}
 {\rm a}_{(j_{1},j_{2})}(M^{2}) = \te \frac{1}{720}(2j_{1}+1)(2j_{2}+1)\big[10\,(j_{1}(j_{1}+1)
 +j_{2}(j_{2}+1))-15\,M^4+60\,M^{2}-58\big]\ , 
 \ee
 so that ${\rm a}(T) = \widehat{\rm a}_{(1,0)}(4)+\widehat{\rm a}_{(0,1)}(4) = -\frac{19}{60}$, 
 in agreement with 
  \cite{Fradkin:1983zz,Beccaria:2014xda}.\footnote{As explained in \cite{Fradkin:1982xc}, 
  it is also possible to express  $T_{\mu\nu}$ in terms of two spin 1 vector fields and thus 
   write the partition function
   in the form (see Eq.(5) of \cite{Fradkin:1983zz}, cf.   (\ref{4.17}))  \ \  $Z_{\rm T}(S^{4}) = 
C \big[\det\widehat\Delta_{1\perp}(3)\big]^{-1}$,
 where $C$  accounts for the  zero mode contributions (cf. discussion  after Eq.(6.18) in 
  \cite{Fradkin:1985am}). 
This zero mode  factor is essential to reproduce the correct value   for the  a-anomaly   giving extra 
 $-\tfrac{1}{2}$ shift: 
 using (\ref{4.37})  to  find  $\widehat{\rm a}_{1\perp}(3)$   we get \ \ \  $\widehat{\rm a}(T) =
2\,\widehat{\rm a}_{1\perp}(3) -\frac{1}{2} = -\frac{19}{60}$.
}
Indeed, the general form of (\ref{4.36}) is \cite{Beccaria:2014jxa}
\be
\begin{split}
\label{E.8}
&{\rm a} (\Delta_{5}; j_{1}, j_{2}) =\te  \frac{1}{720}(2j_{1}+1)(2j_{2}+1)(\Delta_{5}-2)
 \\
& \times \Big[ -3(\Delta_{5}-2)^{4}+10(j_{1}^{2}+j_{2}^{2}+j_{1}+j_{2}+\tfrac{1}{2})(\Delta_{5}-2)^{2}
-15(j_{1}-j_{2})^{2}(j_{1}+j_{2}+1)^{2}
\Big]\ ,
\end{split}
\ee
and we  again  get 
$
{\rm a} (3;1,0)+ {\rm a} (3;0,1)  \te = -\frac{19}{60}$.

One can repeat the above discussion  for  a conformal  4d field $T_p$ transforming in the $(p,0)\oplus(0,p)$ representation of the $SO(1,3)$.\footnote{Fields transforming in the $(p,0)\oplus(0,p)$  representation are 
 Weyl-like tensors  (see,  e.g.,  \cite{Gomez-Avila:2013qaa}).
They may be represented as rank $2p$ tensors $T_{\mu_{1}\nu_{1}\mu_{2}\nu_{2}\cdots\mu_{p}\nu_{p}}$
antisymmetric in each  pair $\mu_{i}\nu_{i}$, totally symmetric   with respect  to 
 the exchange of the pairs $(\mu_{i}\nu_{i})$ and $(\mu_{j}\nu_{j})$, traceless, and
  obeying a generalized "Bianchi identity" 
   $T_{\cdots[\mu\nu\rho]}=0$.
}
We may start in  \ads{5}  with a 5d   field in $(\Delta_{5}; p,0)\oplus(\Delta_{5}; 0,p)$
representation (to be denoted as HS$_p$).  It   should   correspond  to  a conformal field in \ads{4} with  the 
canonical dimension $4-\Delta_{5}$
and thus with  the kinetic term $T_p\Box^{\Delta_{5}-2}T_p+...$. 
The  correspondence rule  (\ref{4.33}) here  reads  as 
\be
\label{E.10}
(\Delta_{5}; p,0)\oplus(\Delta_{5}; 0,p)
\rightarrow  Z_{{\text{T}_p}} ({\rm  AdS_4}) = \prod_{k=1}^{\Delta_{5}-2}
\Big[ \det\widehat\Delta_{(p,0)\oplus(0,p)}\big(2+p(p+1)-k(k-1)\big)\Big]^{-1/2} 
\ee
Using (\ref{E.6}) and (\ref{E.8})  one finds  that 
the equality  of the  a-anomaly coefficients  implied by \rf{E.10} indeed   holds 
\be
\label{E.11}
{\rm a} (\Delta_{5};p,0)+{\rm a} (\Delta_{5};0,p) = 2\,\sum_{k=1}^{\Delta_{5}-1}
{\rm a}_{(p,0)}(2+p(p+1)-k(k-1))\ .
\ee
From  (\ref{E.5}) we find that  the  dimension corresponding according to \rf{2.2} 
 to the $k$-th operator 
 in the r.h.s. of (\ref{E.10}) is   $\Delta_{4}^{(k)} = k+1$, so that from \rf{2.4}   we get 
\be
\begin{split}
\label{E.12}
\Z_{{\rm HS}_p}^{+}(\ads{5}; q) &= \frac{2\,(2p+1)\,q^{\Delta_{5}}}{(1-q)^{4}}, \\
\Z_{\text{T}_p}^{+}(\ads{4}; q) &= \frac{2(2p+1)}{(1-q)^{3}}\sum_{k=1}^{\Delta_{5}-2}q^{k+1} = 
\frac{2(2p+1)}{(1-q)^{4}}(q^{2}-q^{\Delta_{5}})\ .
\end{split}
\ee
Thus once again we get  the relation \rf{3.1a} 
\be
\label{E.13}
\bZ^{-}_{{\rm HS}_p}(\ads{5}; q) - \Z^{+}_{{\rm HS}_p}(\ads{5}; q)  = 
\bZ^{-}_{\text{T}_p}(\ads{4}; q) +\Z^{+}_{\text{T}_p}(\ads{4}; q) \ . 
\ee

\iffa

\section{Papers of interest not yet cited}


we may add \cite{Aharony:2010ay}  as ref -- see p 3 there.
Also \cite{Hinterbichler:2015pta}
may worth looking at -- there may be connections.
Notice  footnote 6 in \cite{Aharony:2010ay}  --  we effectively have
a related observation.
there is also some relevant discussion  in 2nd paragraph below 5.5
so we may split the cases: (1)  theory defined on boundary of \ads{5}, (2)
theory formally defined  on \ads{4} rather than its double  cover.

\noindent
Ref \cite{Ohl:2012bk} discusses "sequential" AdS/CFT  in their  last sect   with
ref to \cite{Compere:2008us,Nilsson:2012ky}
But they refer to CS  type 3d gravity / higher spins    which is not
what we want/get so we should distinguish our discussion from this.
but looks like Nilsson  in \cite{Nilsson:2012ky}   was  first to speak about
$\ads{4}/\text{CF}_{3}\to \ads{3}/\text{CF}_{2}$.

\fi

\bibliography{BT-Biblio}

\providecommand{\href}[2]{#2}\begingroup\raggedright\begin{thebibliography}{10}

\bibitem{Beccaria:2014jxa}
M.~Beccaria, X.~Bekaert, and A.~A. Tseytlin, {\it {Partition function of free
  conformal higher spin theory}},  {\em JHEP} {\bf 1408} (2014) 113,
  [\href{http://arxiv.org/abs/1406.3542}{{\tt arXiv:1406.3542}}].

\bibitem{Giombi:2013yva}
S.~Giombi, I.~R. Klebanov, S.~S. Pufu, B.~R. Safdi, and G.~Tarnopolsky, {\it
  {AdS Description of Induced Higher-Spin Gauge Theory}},  {\em JHEP} {\bf
  1310} (2013) 016, [\href{http://arxiv.org/abs/1306.5242}{{\tt
  arXiv:1306.5242}}].

\bibitem{Beccaria:2014xda}
M.~Beccaria and A.~A. Tseytlin, {\it {Higher spins in AdS$_{5}$ at one loop:
  vacuum energy, boundary conformal anomalies and AdS/CFT}},  {\em JHEP} {\bf
  1411} (2014) 114, [\href{http://arxiv.org/abs/1410.3273}{{\tt
  arXiv:1410.3273}}].

\bibitem{Breitenlohner:1982bm}
P.~Breitenlohner and D.~Z. Freedman, {\it {Positive Energy in anti-De Sitter
  Backgrounds and Gauged Extended Supergravity}},  {\em Phys. Lett.} {\bf B115}
  (1982) 197.

\bibitem{Nilsson:2012ky}
B.~E.~W. Nilsson, {\it {Aspects of topologically gauged M2-branes with six
  supersymmetries: towards a 'sequential AdS/CFT'?}},  in {\em {Quantum theory
  and symmetries. Proceedings, 7th International Conference, QTS7, Prague,
  Czech Republic, August 7-13, 2011}}, 2012.
\newblock \href{http://arxiv.org/abs/1203.5090}{{\tt arXiv:1203.5090}}.

\bibitem{Ohl:2012bk}
T.~Ohl and C.~F. Uhlemann, {\it {Saturating the unitarity bound in
  AdS/CFT$_\text{(AdS)}$}},  {\em JHEP} {\bf 05} (2012) 161,
  [\href{http://arxiv.org/abs/1204.2054}{{\tt arXiv:1204.2054}}].

\bibitem{Karch:2000ct}
A.~Karch and L.~Randall, {\it {Locally localized gravity}},  {\em JHEP} {\bf
  05} (2001) 008, [\href{http://arxiv.org/abs/hep-th/0011156}{{\tt
  hep-th/0011156}}].

\bibitem{Compere:2008us}
G.~Compere and D.~Marolf, {\it {Setting the boundary free in AdS/CFT}},  {\em
  Class. Quant. Grav.} {\bf 25} (2008) 195014,
  [\href{http://arxiv.org/abs/0805.1902}{{\tt arXiv:0805.1902}}].

\bibitem{Aharony:2010ay}
O.~Aharony, D.~Marolf, and M.~Rangamani, {\it {Conformal field theories in
  anti-de Sitter space}},  {\em JHEP} {\bf 02} (2011) 041,
  [\href{http://arxiv.org/abs/1011.6144}{{\tt arXiv:1011.6144}}].

\bibitem{Andrade:2011nh}
T.~Andrade and C.~F. Uhlemann, {\it {Beyond the unitarity bound in
  AdS/CFT$_\text{(A)dS}$}},  {\em JHEP} {\bf 01} (2012) 123,
  [\href{http://arxiv.org/abs/1111.2553}{{\tt arXiv:1111.2553}}].

\bibitem{Hinterbichler:2015pta}
K.~Hinterbichler, J.~Stokes, and M.~Trodden, {\it {Holographic CFTs on
  maximally symmetric spaces: correlators, integral transforms and
  applications}},  {\em Phys. Rev.} {\bf D92} (2015), no.~6 065025,
  [\href{http://arxiv.org/abs/1505.05513}{{\tt arXiv:1505.05513}}].

\bibitem{Metsaev:2000qb}
R.~R. Metsaev, {\it {Massive fields in AdS(3) and compactification in AdS space
  time}},  {\em Nucl. Phys. Proc. Suppl.} {\bf 102} (2001) 100--106,
  [\href{http://arxiv.org/abs/hep-th/0103088}{{\tt hep-th/0103088}}].
  [,100(2000)].

\bibitem{Artsukevich:2008vy}
A.~{\relax Yu}. Artsukevich and M.~A. Vasiliev, {\it {On Dimensional Degression
  in AdS(d)}},  {\em Phys. Rev.} {\bf D79} (2009) 045007,
  [\href{http://arxiv.org/abs/0810.2065}{{\tt arXiv:0810.2065}}].

\bibitem{Dowker:1983nt}
J.~S. Dowker, {\it {Arbitrary Spin Theory in the Einstein Universe}},  {\em
  Phys. Rev.} {\bf D28} (1983) 3013.

\bibitem{Hawking:1999dp}
S.~W. Hawking and H.~S. Reall, {\it {Charged and rotating AdS black holes and
  their CFT duals}},  {\em Phys. Rev.} {\bf D61} (2000) 024014,
  [\href{http://arxiv.org/abs/hep-th/9908109}{{\tt hep-th/9908109}}].

\bibitem{Metsaev:2007rw}
R.~R. Metsaev, {\it {Ordinary-derivative formulation of conformal totally
  symmetric arbitrary spin bosonic fields}},  {\em JHEP} {\bf 1206} (2012) 062,
  [\href{http://arxiv.org/abs/0709.4392}{{\tt arXiv:0709.4392}}].

\bibitem{Metsaev:1994ys}
R.~R. Metsaev, {\it {Lowest eigenvalues of the energy operator for totally
  (anti)symmetric massless fields of the n-dimensional anti-de Sitter group}},
  {\em Class.Quant.Grav.} {\bf 11} (1994) L141--L145.

\bibitem{Metsaev:1995re}
R.~R. Metsaev, {\it {Massless mixed symmetry bosonic free fields in
  d-dimensional anti-de Sitter space-time}},  {\em Phys.Lett.} {\bf B354}
  (1995) 78--84.

\bibitem{Metsaev:2003cu}
R.~R. Metsaev, {\it {Massive totally symmetric fields in AdS$_{d}$}},  {\em
  Phys.Lett.} {\bf B590} (2004) 95--104,
  [\href{http://arxiv.org/abs/hep-th/0312297}{{\tt hep-th/0312297}}].

\bibitem{Tseytlin:2013jya}
A.~A. Tseytlin, {\it {On partition function and Weyl anomaly of conformal
  higher spin fields}},  {\em Nucl.Phys.} {\bf B877} (2013) 598--631,
  [\href{http://arxiv.org/abs/1309.0785}{{\tt arXiv:1309.0785}}].

\bibitem{Tseytlin:2013fca}
A.~A. Tseytlin, {\it {Weyl anomaly of conformal higher spins on six-sphere}},
  {\em Nucl.Phys.} {\bf B877} (2013) 632--646,
  [\href{http://arxiv.org/abs/1310.1795}{{\tt arXiv:1310.1795}}].

\bibitem{Irakleidou:2015exd}
M.~Irakleidou and I.~Lovrekovic, {\it {Conformal gravity 1-loop partition
  function}},  \href{http://arxiv.org/abs/1512.07130}{{\tt arXiv:1512.07130}}.

\bibitem{Lovrekovic:2015thw}
I.~Lovrekovic, {\it {One loop partition function of six dimensional conformal
  gravity using heat kernel on AdS}},
  \href{http://arxiv.org/abs/1512.00858}{{\tt arXiv:1512.00858}}.

\bibitem{Gopakumar:2011qs}
R.~Gopakumar, R.~K. Gupta, and S.~Lal, {\it {The Heat Kernel on $AdS$}},  {\em
  JHEP} {\bf 1111} (2011) 010, [\href{http://arxiv.org/abs/1103.3627}{{\tt
  arXiv:1103.3627}}].

\bibitem{Gupta:2012he}
R.~K. Gupta and S.~Lal, {\it {Partition Functions for Higher-Spin theories in
  AdS}},  {\em JHEP} {\bf 1207} (2012) 071,
  [\href{http://arxiv.org/abs/1205.1130}{{\tt arXiv:1205.1130}}].

\bibitem{Giombi:2014yra}
S.~Giombi, I.~R. Klebanov, and A.~A. Tseytlin, {\it {Partition Functions and
  Casimir Energies in Higher Spin $AdS_{d+1}/CFT_d$}},  {\em Phys. Rev.} {\bf
  D90} (2014), no.~2 024048, [\href{http://arxiv.org/abs/1402.5396}{{\tt
  arXiv:1402.5396}}].

\bibitem{Erdmenger:1997wy}
J.~Erdmenger and H.~Osborn, {\it {Conformally covariant differential operators:
  Symmetric tensor fields}},  {\em Class.Quant.Grav.} {\bf 15} (1998) 273--280,
  [\href{http://arxiv.org/abs/gr-qc/9708040}{{\tt gr-qc/9708040}}].

\bibitem{Beccaria:2015vaa}
M.~Beccaria and A.~Tseytlin, {\it {On higher spin partition functions}},  {\em
  J.Phys.} {\bf A48} (2015), no.~27 275401,
  [\href{http://arxiv.org/abs/1503.08143}{{\tt arXiv:1503.08143}}].

\bibitem{Fradkin:1985am}
E.~S. Fradkin and A.~A. Tseytlin, {\it {Conformal supergravity}},  {\em
  Phys.Rept.} {\bf 119} (1985) 233--362.

\bibitem{Vasiliev:2009ck}
M.~A. Vasiliev, {\it {Bosonic conformal higher-spin fields of any symmetry}},
  {\em Nucl. Phys.} {\bf B829} (2010) 176--224,
  [\href{http://arxiv.org/abs/0909.5226}{{\tt arXiv:0909.5226}}].

\bibitem{Bekaert:2013zya}
X.~Bekaert and M.~Grigoriev, {\it {Higher order singletons, partially massless
  fields and their boundary values in the ambient approach}},  {\em Nucl.Phys.}
  {\bf B876} (2013) 667--714, [\href{http://arxiv.org/abs/1305.0162}{{\tt
  arXiv:1305.0162}}].

\bibitem{Barnich:2015tma}
G.~Barnich, X.~Bekaert, and M.~Grigoriev, {\it {Notes on conformal invariance
  of gauge fields}},  {\em J. Phys.} {\bf A48} (2015), no.~50 505402,
  [\href{http://arxiv.org/abs/1506.00595}{{\tt arXiv:1506.00595}}].

\bibitem{Beccaria:2015uta}
M.~Beccaria and A.~A. Tseytlin, {\it {Conformal a-anomaly of some non-unitary
  6d superconformal theories}},  {\em JHEP} {\bf 09} (2015) 017,
  [\href{http://arxiv.org/abs/1506.08727}{{\tt arXiv:1506.08727}}].

\bibitem{Dolan:2005wy}
F.~Dolan, {\it {Character formulae and partition functions in higher
  dimensional conformal field theory}},  {\em J.Math.Phys.} {\bf 47} (2006)
  062303, [\href{http://arxiv.org/abs/hep-th/0508031}{{\tt hep-th/0508031}}].

\bibitem{Metsaev:1995jp}
R.~R. Metsaev, {\it {All conformal invariant representations of d-dimensional
  anti-de Sitter group}},  {\em Mod. Phys. Lett.} {\bf A10} (1995) 1719--1731.

\bibitem{Metsaev:2008ba}
R.~Metsaev, {\it {Conformal self-dual fields}},  {\em J.Phys.} {\bf A43} (2010)
  115401, [\href{http://arxiv.org/abs/0812.2861}{{\tt arXiv:0812.2861}}].

\bibitem{Tseytlin:2002gz}
A.~A. Tseytlin, {\it {On limits of superstring in $AdS_{5}\times S^{5}$}},
  {\em Theor.Math.Phys.} {\bf 133} (2002) 1376--1389,
  [\href{http://arxiv.org/abs/hep-th/0201112}{{\tt hep-th/0201112}}].

\bibitem{Osborn:1993cr}
H.~Osborn and A.~C. Petkou, {\it {Implications of conformal invariance in field
  theories for general dimensions}},  {\em Annals Phys.} {\bf 231} (1994)
  311--362, [\href{http://arxiv.org/abs/hep-th/9307010}{{\tt hep-th/9307010}}].

\bibitem{Leigh:2003ez}
R.~G. Leigh and A.~C. Petkou, {\it {SL(2,Z) action on three-dimensional CFTs
  and holography}},  {\em JHEP} {\bf 12} (2003) 020,
  [\href{http://arxiv.org/abs/hep-th/0309177}{{\tt hep-th/0309177}}].

\bibitem{Pope:1989vj}
C.~N. Pope and P.~K. Townsend, {\it {Conformal Higher Spin in
  (2+1)-dimensions}},  {\em Phys. Lett.} {\bf B225} (1989) 245.

\bibitem{Fradkin:1989xt}
E.~S. Fradkin and V.~{\relax Ya}. Linetsky, {\it {A Superconformal Theory of
  Massless Higher Spin Fields in $D$ = (2+1)}},  {\em Mod. Phys. Lett.} {\bf
  A4} (1989) 731. [Annals Phys.198,293(1990)].

\bibitem{Nilsson:2015pua}
B.~E.~W. Nilsson, {\it {On the conformal higher spin unfolding equation for a
  three-dimensional self-interacting scalar field}},
  \href{http://arxiv.org/abs/1506.03328}{{\tt arXiv:1506.03328}}.

\bibitem{Deser:1981wh}
S.~Deser, R.~Jackiw, and S.~Templeton, {\it {Topologically Massive Gauge
  Theories}},  {\em Annals Phys.} {\bf 140} (1982) 372--411. [Annals
  Phys.281,409(2000)].

\bibitem{Horne:1988jf}
J.~H. Horne and E.~Witten, {\it {Conformal Gravity in Three-dimensions as a
  Gauge Theory}},  {\em Phys. Rev. Lett.} {\bf 62} (1989) 501--504.

\bibitem{Afshar:2011qw}
H.~Afshar, B.~Cvetkovic, S.~Ertl, D.~Grumiller, and N.~Johansson, {\it
  {Conformal Chern-Simons holography - lock, stock and barrel}},  {\em Phys.
  Rev.} {\bf D85} (2012) 064033, [\href{http://arxiv.org/abs/1110.5644}{{\tt
  arXiv:1110.5644}}].

\bibitem{Witten:2003ya}
E.~Witten, {\it {SL(2,Z) action on three-dimensional conformal field theories
  with Abelian symmetry}},  \href{http://arxiv.org/abs/hep-th/0307041}{{\tt
  hep-th/0307041}}.

\bibitem{Tseytlin:1984wj}
A.~A. Tseytlin, {\it {Effective action in de Sitter space and conformal
  supergravity}},  {\em Yad.Fiz.} {\bf 39} (1984), no.~6 1606--1615.

\bibitem{Fradkin:1983zz}
E.~S. Fradkin and A.~A. Tseytlin, {\it {Instanton zero modes and beta functions
  in supergravities. 2. Conformal supergravity}},  {\em Phys.Lett.} {\bf B134}
  (1984) 307.

\bibitem{Fradkin:1981jc}
E.~S. Fradkin and A.~A. Tseytlin, {\it {One Loop Beta Function in Conformal
  Supergravities}},  {\em Nucl.Phys.} {\bf B203} (1982) 157.

\bibitem{Deser:1983mm}
S.~Deser and R.~I. Nepomechie, {\it {Gauge Invariance Versus Masslessness in De
  Sitter Space}},  {\em Annals Phys.} {\bf 154} (1984) 396.

\bibitem{Juhl:2011aa}
A.~{Juhl}, {\it {Explicit formulas for GJMS-operators and \$Q\$-curvatures}},
  {\em ArXiv e-prints} (Aug., 2011) [\href{http://arxiv.org/abs/1108.0273}{{\tt
  arXiv:1108.0273}}].

\bibitem{Gover:2005mn}
A.~R. Gover, {\it {Laplacian operators and Q-curvature on conformally Einstein
  manifolds}},  \href{http://arxiv.org/abs/math/0506037}{{\tt math/0506037}}.

\bibitem{Juhl:2009aa}
A.~{Juhl}, {\it {On conformally covariant powers of the Laplacian}},  {\em
  ArXiv e-prints} (May, 2009) [\href{http://arxiv.org/abs/0905.3992}{{\tt
  arXiv:0905.3992}}].

\bibitem{Paneitz:1983}
S.~Paneitz, {\it {A Quartic Conformally Covariant Differential Operator for
  Arbitrary Pseudo-Riemannian Manifolds (Summary)}},
  \href{http://arxiv.org/abs/0803.4331}{{\tt arXiv:0803.4331}}.

\bibitem{Christensen:1978md}
S.~Christensen and M.~Duff, {\it {New Gravitational Index Theorems and
  Supertheorems}},  {\em Nucl.Phys.} {\bf B154} (1979) 301.

\bibitem{Fradkin:1982xc}
E.~S. Fradkin and A.~A. Tseytlin, {\it {Asymptotic freedom in extended
  conformal supergravities}},  {\em Phys. Lett.} {\bf B110} (1982) 117--122.

\bibitem{Gomez-Avila:2013qaa}
S.~G\'omez-\'Avila and M.~Napsuciale, {\it {Covariant basis induced by parity
  for the $(j,0)\oplus (0,j)$ representation}},  {\em Phys. Rev.} {\bf D88}
  (2013), no.~9 096012, [\href{http://arxiv.org/abs/1307.4711}{{\tt
  arXiv:1307.4711}}].

\end{thebibliography}\endgroup
\bibliographystyle{JHEP}

\end{document}